# Quantifying the Benefit of Artificial Intelligence for Scientific Research

Jian Gao[1,2,3], Dashun Wang[1,2,3,4*]

[1] Center for Science of Science and Innovation, Northwestern University, Evanston, IL, 60208, USA
[2] Kellogg School of Management, Northwestern University, Evanston, IL, 60208, USA
[3] Northwestern Institute on Complex Systems, Northwestern University, Evanston, IL, 60208, USA
[4] McCormick School of Engineering, Northwestern University, Evanston, IL, 60208, USA

* Correspondence to: dashun.wang@northwestern.edu

**The ongoing artificial intelligence (AI) revolution has the potential to change almost every line of work. As AI capabilities continue to improve in accuracy, robustness, and reach, AI may outperform and even replace human experts across many valuable tasks. Despite enormous effort devoted to understanding the impact of AI on labor and the economy and AI's recent successes in accelerating scientific discovery and progress, we lack a systematic understanding of how AI advances may benefit scientific research across disciplines and fields. Here, drawing from the literature on the future of work and the science of science, we develop a measurement framework to estimate both the direct use of AI and the potential benefit of AI in scientific research, applying natural language processing techniques to 74.6 million publications and 7.1 million patents. We find that the use of AI in research is widespread throughout the sciences, growing especially rapidly since 2015, and papers that use AI exhibit a citation premium, more likely to be highly cited both within and outside their disciplines. Moreover, our analysis reveals considerable potential for AI to benefit numerous scientific fields, yet a notable disconnect exists between AI education and its research applications, highlighting a mismatch between the supply of AI expertise and its demand in research. Lastly, we examine demographic disparities in AI's benefits across scientific disciplines and find that disciplines with a higher proportion of women or Black scientists tend to be associated with less benefit, suggesting that AI's growing impact on research may further exacerbate existing inequalities in science. As the connection between AI and scientific research deepens, our findings may become increasingly important, with implications for the equity and sustainability of the research enterprise.**

# Main text

The rapid advances in artificial intelligence (AI) may lead to massive value creation and capture across many facets of human society[1-5], creating enormous social and economic opportunities[6-8], and just as many challenges[9-17]. While extensive efforts have been devoted to understanding the impact of AI on the labor market and the economy[18-22], the impact of AI on the growing research enterprise remains unclear. Indeed, recent AI advances have shown promise to achieve and, in some cases, exceed expert-level performance across many economically valuable tasks[23-30]. As society prepares for the moment when AI may outperform or even replace human recruiters, bankers, doctors, lawyers, composers, and drivers, an important question arises: What is the impact of AI in advancing scientific research across different disciplines and fields?

A better understanding of the impact of AI on science may not only help guide AI development, bridging AI innovations more closely with scientific research but also hold implications for science and innovation policy. Understanding the impact of AI on science is both timely and important given AI's recent remarkable success in advancing research frontiers across several fields[31-44], from predicting the structure of proteins in biology[45-47] to designing new drug candidates in medicine[48-51], from discovering natural laws in physics[52-54] to solving complicated equations and discovering new conjectures in mathematics[55-57], from controlling nuclear fusion[58] to predicting new material properties[59-62], from designing taxation policy[63] to suggesting democratic social mechanisms[64], and many more[65-70]. These advances raise the possibility that as AI continues to improve in accuracy, robustness, and reach[71-76], it may bring meaningful benefits to science, propelling scientific progress across a range of research areas while significantly augmenting researchers' innovative capacities.

Yet, despite the rapid progress of AI and its broad applications in several domains, there is substantial skepticism about whether today's AI is capable or significant enough to advance scientific research. Indeed, most current AI applications belong to the category of "narrow AI,"[77-79] which tackles specifically defined problems, and hence may not be suitable to fulfill the broad range of tasks that scientific research demands[2,18,80]. Further, to the extent that AI may provide automated solutions to an existing problem, science is about not only solving well-defined problems but also spotting new frontiers and generating novel hypotheses[81]. These views paint a more nuanced picture of AI's applicability to advancing science, suggesting that AI may be better suited to perform some research tasks than others[2,10,82].

Building on the growing literature on the future of work[83-87] and the science of science[88-96], here we develop a quantitative framework for estimating the impact of AI on scientific research (see Methods for details). Our primary dataset contains 74.6 million publications from 1960 to 2019 from the Microsoft Academic Graph (MAG) dataset[97], spanning 19 disciplines and 292 fields (see Supplementary Note 1.1 for details). We integrate this dataset with 7.1 million patents granted between 1976 and 2019 by the U.S. Patent and Trademark Office (USPTO) (see Supplementary Note 1.2). We then follow prior studies to identify AI publications and AI patents using a keyword-based approach (see Supplementary Notes 2.1 and 3.1 for details)[91,97,98], allowing us to measure AI's impact on science at two levels. First, we quantify the direct impact of AI using an "AI n-gram framework" (**Fig. 1a**), which estimates the relative frequency of the use of AI in a field (see Supplementary Notes 2.3). Specifically, we extract AI n-grams (bigrams and trigrams; e.g., "deep learning" and "convolutional neural network") from both the titles and abstracts of AI publications and calculate the frequency of their occurrences to approximate AI advances[91,96]. We then repeat this n-gram measurement for publications in each field and year, allowing us to calculate the weighted frequency of AI n-grams appearing in a paper to approximate the direct impact of AI on each field and year. Second, motivated by the future of work literature[82-84], we quantify the potential impact of AI using an "AI capability–field task framework" (**Fig. 2a**), which measures the alignment between AI capabilities and the tasks of a field (see Supplementary Notes 3.3). In particular, we infer the capabilities of AI (i.e., what AI can do) by extracting verb-noun pairs (e.g., "learn representation") from the titles of AI publications and AI patents using natural language processing (NLP) techniques and calculating their relative frequency[99-101]. Here, following prior work[82], we rely only on titles as they have a higher signal-to-noise ratio than abstracts. We then estimate the tasks of each field (i.e., what a field does) by calculating the relative frequency of verb-noun pairs extracted from the titles of publications in each field and year. Calculating the overlap between the prevalent tasks in a field and the inferred AI capabilities allows us to approximate the potential impact of AI on each field and year (see Methods for more details).

## Results

**The widespread use of AI across the sciences**

Overall, AI research presents a dynamically evolving landscape (**Figs. 1b,c**). While the frequency of certain dominant AI n-grams in 2019 (e.g., "machine learning," "convolutional neural network," "deep learning," "deep neural network," and "artificial intelligence") shows an overall upward trend (**Fig. 1b**), some AI n-grams emerged only recently (e.g., "generative adversarial network"), some rose to prominence after a long period of dormancy (e.g., "deep learning"), and some were popular a decade ago but have become less prevalent in recent years (e.g., "support vector machine"; **Fig. 1c**). Amidst this rapidly evolving AI research landscape, there has been a precipitous rise in the use of AI by many disciplines, as proxied by the mention of AI n-grams in the titles and abstracts of publications (**Fig. 1d**; see Supplementary Fig. 2 for details).

This increase in the use of AI by different disciplines raises an interesting question: How do the citations of papers that use AI compare to those of other papers in the same field? To answer this question, we define hit papers as those in the top 5% by total citations in the same field and year and calculate the likelihood that a paper is a hit paper. We find that for a majority of disciplines, papers that mention AI n-grams tend to be associated with a higher probability of being hit papers within their disciplines (**Fig. 1e**), and they also receive more citations from other disciplines compared to papers that do not mention AI n-grams (**Fig. 1f**). This citation premium of papers that mention AI n-grams appears to be stronger in disciplines with a lower propensity to use AI (see detailed correlation analysis in Supplementary Fig. 4), suggesting that disciplines that seem distant from AI may see substantial benefits from using AI to advance their research.

The dynamic AI landscape also prompts us to explore changes in the direct impact of AI on scientific research over time. Specifically, we calculate the direct AI impact score using the "AI n-gram framework" for each discipline between 2000 and 2019 (see Supplementary Fig. 3 for the results for 1960-2019), extracting AI n-grams from the titles and abstracts of AI publications and calculating their frequency of occurrence in the publications in each discipline (see Methods and Supplementary Note 2 for details). We find that disciplines overall used more AI in their research over the past two decades (e.g., in computer science, the direct AI impact score increased from 0.5% in 2000 to 1.3% in 2019; **Fig. 1g**, solid lines). This increase occurs not only in computer science but also in a wide range of other disciplines (**Fig. 1h**), including, for example, physics, biology, and economics. Moreover, this increase has not been linear; there has been a notably sharp increase in the use of AI since 2015 across many disciplines (**Fig. 1g**).

To better understand whether the recent rise in the direct impact of AI on science is associated with changes in AI capabilities or field-specific shifts in research direction, we calculate an alternative measure for AI impact scores by keeping AI capabilities fixed in 2015 and apply this alternative measure to each discipline and year between 2015 and 2019 (see Supplementary Note 4.2). More specifically, we use (1) AI n-grams extracted from AI publications before 2015 without changing either their terms or their frequency for 2015-2019, and (2) AI n-grams extracted from publications in each discipline and year during the period 2015-2019 (**Fig. 1g**, dashed lines). We find that the new scores deviate substantially from the original scores (**Fig. 1g**, solid lines, and Supplementary Fig. 7), which indicates that across disciplines, sciences benefit more from cutting-edge AI advances. Overall, these results suggest that new AI capabilities play an important role in contributing to the recent, sharp increase in the use of AI across scientific disciplines (see Supplementary Note 4 for more details and results related to AI's growing impact).

**The potential impact of AI**

While explicit mentions of AI n-grams by publications signal the direct use of AI in research, AI may also exert potential impacts on scientific research beyond these direct uses. In particular, the growing AI capabilities may help perform some core tasks that a research field demands. Here we build on the future of work literature, which suggests that AI capabilities and field tasks can be captured by verb-noun pairs (e.g., "learn representation")[82-84], prompting us to develop an "AI capability–field task framework" to quantify the potential impact of AI on scientific research (**Fig. 2a**). We apply NLP algorithms to extract verb-noun pairs from the titles of AI papers and AI patents to estimate AI capabilities[99-101] (see Methods for details and Supplementary Note 8.3 for validations of the approach). Applied systematically to all disciplines and fields, this framework allows us to estimate which subfields within a discipline may benefit most from AI. Taking biology as an example, the results suggest that the subfield with the largest potential AI impact is "biological system" (**Fig. 2c**), as many of its basic tasks appear aligned with inferred AI capabilities (e.g., "extract feature," "detect object," and "improve prediction"). Interestingly, the "biological system" field, ranked seventh among all non-computer science fields for the potential AI impact (**Fig. 2d**), also happens to be the field for the AlphaFold paper[45], which *Science* called the 2021 Breakthrough of the Year[102].

While there are considerable differences in the direct impact of AI on scientific disciplines (**Fig. 1h**), the differences in the potential AI impact are relatively small across disciplines (Supplementary Fig.7), suggesting the potentially widespread applicability of AI in science. We further study within-discipline heterogeneity by examining the percentiles of the direct and potential AI impact scores for each discipline's subfields (see Supplementary Note 4.1 for details). We find that the two percentiles in each field are highly correlated with each other (**Fig. 2e**; Pearson's $r$ = 0.891 and $P$-value < 0.001), especially among the top-ranked subfields (**Fig. 2f**; the top 3 subfields within each discipline according to the two percentiles are entirely overlapped in over half disciplines), indicating that the two measurements are consistent in identifying fields most benefit from AI (see Supplementary Table 1 for the list of three subfields within each discipline that have the highest direct AI impact scores and potential AI impact scores). Nevertheless, almost every discipline has some subfields with a significant AI impact, which holds robust even for disciplines with a low AI impact overall, such as sociology and economics (**Fig. 2g**; see Supplementary Fig. 8 for the results for all disciplines). Taken together, these results suggest that the direct impact of AI on research is pervasive across disciplines and fields, and its potential impact may extend beyond its current uses in science.

**Growing knowledge demands for AI**

The rapidly expanding AI frontier and its increasing impact on science may lead to growing demands for AI expertise from domain experts, raising the question of whether the current education and training on AI skills are commensurate with AI's impact. To answer this question, we analyze 4.2 million university course syllabi from the Open Syllabus Project (OSP) database[103] and estimate the level of AI education in each discipline (see Methods and Supplementary Note 5 for details). We find that, excluding the top three computational disciplines (i.e., computer science, mathematics, and engineering), the correlation between the AI education level in a discipline and the impact of AI on the discipline decreases, as well as its significance (**Figs. 3a,b**; Pearson's $r$ = 0.493 with $P$-value = 0.074 for direct impact and Pearson's $r$ = 0.263 with $P$-value = 0.363 for potential impact). The results suggest that the supply of AI talent and knowledge in most disciplines appears to be incommensurate with the benefits these disciplines may extract from AI capabilities, highlighting a substantial AI impact–AI training gap. This result is robust under alternative measures of AI education levels (Supplementary Fig. 12).

To meet the growing knowledge demands on AI, domain experts may rely on cross-discipline collaborations to access AI capabilities. We analyze collaboration patterns for AI publications in domains other than computer science that are co-authored by domain experts and/or computer scientists (as a proxy for AI researchers; see Methods and Supplementary Notes 6.1 and 6.2 for detailed methods and alternative proxies of AI researchers). We find that, in aggregate, about 40% of AI publications are published by domain experts and about one-third are collaborative works (**Fig. 3c**). In disciplines where AI has a greater impact, we see a larger propensity for domain experts to collaborate with computer scientists (**Figs. 3d,e**; Pearson's $r$ = 0.841 and $P$-value < 0.001 for direct impact; Pearson's $r$ = 0.802 and $P$-value < 0.001 for potential impact). Moreover, we see faster growth in collaborative AI publications than in domain-only AI publications (Supplementary Fig. 13). Overall, the share of collaborative AI publications in major disciplines is increasing over time (**Fig. 3f**; see Supplementary Fig. 14 for the results for other disciplines), suggesting that domain experts' reliance on AI expertise is growing. These results are robust when we use an alternative method of determining AI researchers (see Supplementary Note 6.3 for detailed methods and results).

We further explored discrepancies between the potential and direct AI impact scores in different fields, as well as temporal changes in AI impacts on a field. More specifically, we dig deeper into fields with "higher-direct-impact" than typical and those with "higher-potential-impact" than typical. We also compare fields that underwent "upshifts" in AI benefits, going from "low-AI-benefit" to "high-AI-benefit" over time. For both cases, we find that those "higher-direct-impact" fields and those fields that experienced "upshifts" in AI benefits tend to feature more cross-domain collaborations with AI researchers and higher education levels in AI (see Supplementary Notes 4.3 for details on measures and statistics). Taken together, these findings highlight the importance of teamwork and cross-domain collaborations amidst AI's potentially increasing impact on scientific research and the narrowing of individual domain expertise across the sciences[104-106].

**Demographic disparities**

As the connection between AI and scientific research deepens, it is important to understand who benefits from AI, which has implications for the equity and sustainability of the research enterprise. Here we study the gender and racial/ethnic composition of each discipline and further examine potential differences in the distribution of the benefits from AI across demographic groups. Specifically, we leverage the de-identified Survey of Doctorate Recipients (SDR) data to solicit demographic information on U.S.-trained doctoral scientists and engineers by the discipline of doctorate, sex, and race/ethnicity. We then crosswalk the SDR disciplines of doctorate to the disciplines in the MAG data to estimate the share of women scientists and underrepresented minorities (URM) scientists in each discipline (see Methods and Supplementary Note 7 for details).

We find a negative correlation between the share of women scientists within each discipline and its AI impact scores for both the direct impact (Pearson's $r = -0.555$; $P$-value = 0.032; **Fig. 4a**) and the potential impact (Pearson's $r = -0.593$; $P$-value = 0.020; **Fig. 4b**). Aggregating the AI impact scores of all disciplines by their gender composition (see Methods for details), we find that women scientists tend to be associated with a smaller score and thus less benefit from AI (**Figs.**

**4cd**). Studying the racial and ethnic composition across disciplines, we find another negative relationship between the share of URM scientists in each discipline and its AI impact score, a pattern that is again robust for both the direct impact of AI (Pearson's $r = -0.734$; $P$-value $= 0.002$; **Fig. 4e**) and its potential impact (Pearson's $r = -0.711$; $P$-value $= 0.003$; **Fig. 4f**). This pattern appears especially strong for Black scientists within the URM group (**Figs. 4gh**). On average, women and URM researchers benefit less from AI. We further performed career-level analysis, looking at what happens when one starts to engage in AI research (see Supplementary Note 9 for details). We find that while the average hit rate of a researcher's papers tends to increase immediately after engaging in AI research, this citation premium is less concentrated among underrepresented groups, with women and URM researchers appearing to profit less from AI engagement compared with their counterparts. Together, these results suggest that while AI has the potential to bring benefits to all disciplines, the benefits may be distributed unequally across demographic groups. Hence as AI's impact on science continues to grow, these unequal career effects may further amplify existing disparities in science[107,108].

## Discussion

In this study, we develop a measurement framework to estimate the extent to which AI may benefit scientific research, aiming to quantify both the direct impact of AI and the potential impact of AI across a range of scientific disciplines and research fields. Here we use impact to broadly represent the benefits of AI to science. We find that scientific disciplines are increasingly using AI, as proxied by the mention of AI-related terms in publication titles and abstracts, with especially sharp growth in recent years. Publications that use AI tend to see a citation premium, as they are more likely to be cited both within and outside their disciplines. While there is substantial heterogeneity in the impact of AI across different disciplines, almost every discipline includes some subfields that see great benefits from AI. For example, the medicine discipline as a whole is not ranked among the highest in terms of AI benefits, but some of its subfields (e.g., "nuclear medicine," "optometry," and "medical physics") show substantial AI benefits (Supplementary Fig. 8). Overall, these results suggest that the benefits that AI may bring to scientific research are widespread across a range of disciplines and fields, potentially extending beyond the current uses of AI in science.

A systematic understanding of the impact of AI on scientific research may better inform science and education policy. Our research suggests that the impact of AI on scientific disciplines has raced ahead across science, facilitated in part by cross-discipline collaborations, while the educational focus on AI to upskill future scientists within each discipline has lagged. This misalignment between the impact of AI and AI education (i.e., the AI impact–AI training gap) has important implications for best practices in preparing next-generation scientists to fully leverage AI advances. While these analyses are correlational by nature, they support the hypothesis that collaboration between domain experts and AI researchers may represent an important way to facilitate the use of AI across science. They also suggest a further benefit of increasing AI training across disciplines, which would likely help the disciplines to develop domain-specific AI expertise, allowing them to enjoy greater and timelier benefits from AI advances.

It is also important to recognize that as AI becomes increasingly capable of performing research tasks, it may have an unequal impact on the research workforce. There are long-standing concerns about demographic disparities in science[109-111]. Our results suggest that the groups that have been historically underrepresented in science are also the groups that may benefit less from AI in scientific research. These results are somewhat expected, given that gender disparities tend to correlate with technical fields, which tend to be dominated by men[112,113]. Nevertheless, our analysis highlights that as AI plays more important roles in accelerating science, it may exacerbate existing disparities in science, with implications for building a diverse, equitable, and inclusive research workforce. It thus underscores the importance of expanding the AI-related professoriate by broadening participation and opportunities in AI research and increasing funding and educational programs targeted toward women and underrepresented groups in AI-related fields[114].

While this study takes an initial step toward quantifying the impact of AI on scientific research, it has several limitations that are important to consider when interpreting the results. First, our analyses build on the future of work literature and rely on publication and patent data. Given its multidimensional nature, however, the benefits of AI for science may go beyond the advantages that can be estimated from such datasets. These frameworks, in fact, may underestimate the full range of benefits that AI may bring to research. AI may, for example, optimize the research process by powering new tools and systems that improve the efficiency of doing science, including improving access to information, reducing the knowledge burden, guiding human intuition, automating routine research tasks, and more[115,116]. Second, AI research evolves rapidly, suggesting the need for continuous monitoring and updates to the estimates of its benefits to science. As our datasets trace publications and patents to the end of 2019, they cannot fully capture newer developments, such as the recent rise of foundation models in AI research[117-120]. Given that these foundation models, such as large language models, can be adapted to a wide range of downstream tasks through fine-tuning, it is conceivable that they may play a significant role in augmenting research. Third, as a general-purpose technology[71,121], AI may generate downstream spillover effects, with indirect impacts on various domains. For example, by discovering faster matrix multiplication algorithms[122], AI may have indirect effects on disciplines that would benefit from such advances. Fourth, although the direct mention of AI n-grams in publication titles and abstracts is suggestive of the use of AI in research, the same n-gram may have different meanings in different contexts. Also, the same AI capability may bring different benefits to different fields, amidst alternative ways to define AI terms (see Supplementary Note 8.1 for details)[63], suggesting fruitful future directions to further improve our frameworks for understanding AI capabilities and their impacts on scientific research. Lastly, as AI's capabilities and its impact on science continue to grow, it will become ever more crucial to understand the impact of AI on fairness and equity in research[16,123]. Equally important is to understand how AI may introduce potential biases or otherwise create unintended consequences in the genesis of scientific knowledge, especially given the "black box" nature of many leading AI tools[124-127].

Overall, these findings based on large-scale quantitative analyses may prove useful to the AI research community, helping us better understand the AI capabilities that may be most fruitful for scientific research. At the same time, the misalignment between the level of AI education and its impact on research suggests that collaborations between domain experts and AI researchers may be especially productive, bridging deep domain expertise and new AI advances. Given that tomorrow's technological developments often begin upstream from basic scientific research[128-130], a more robust understanding of the impact of AI on science may further inform a range of important policy considerations for the future of education, research, and innovation[2-4].

## Methods

**Data sources.** To estimate the impact of AI on science, we use a variety of datasets that include information regarding scientific publications, patents, course syllabi, and the demographics of researchers (see Supplementary Note 1 for details). We introduce two primary datasets. (1) We use the Microsoft Academic Graph (MAG) database for publication data. We collect information on 74.6 million publications between 1960 and 2019 of various types ("journal," "conference," "book," or "book chapter"). These publications are categorized into 19 disciplines (e.g., "computer science") and 292 fields (e.g., "machine learning") under the MAG "field of study" taxonomy, in which one discipline contains several child fields (see Supplementary Note 1.1 for details). For each publication, we collect the title, abstract, year, discipline, and field information. (2) We use PatentsView for patent data. We collect information on 7.1 million patents granted between 1976 and 2019 from PatentsView, a data platform based on bulk data from the U.S. Patent and Trademark Office (USPTO). Each patent is associated with a list of patent classification codes and keywords. Using these codes and keywords, we identify AI-related patents (see Supplementary Note 1.2 for details). Together, the MAG publication data and USPTO patent data allow us to estimate the direct and potential impacts of AI on disciplines and fields.

We supplement the analysis with two more datasets to examine the alignment of the impact of AI with the level of AI education and to study the gender, racial, and ethnic composition in science. (1) We use syllabus data that is sourced from the Open Syllabus Project (OSP), the world's first large-scale database of university course syllabus documents. Our syllabus dataset contains 4.2 million English-language syllabi published between 2000 and 2018 (see Supplementary Note 1.3 for details). Each syllabus is associated with a list of Classification of Instructional Programs (CIP) codes representing its academic fields and a list of referenced publications[103]. We manually crosswalk CIP codes to MAG fields, and we link syllabus references to MAG publications (see Supplementary Note 5.1 for details). (2) We use the Survey of Doctorate Recipients (SDR) for de-identified demographic data regarding individuals with a U.S. research doctoral degree in a science, engineering, or health field. We use the 2017 SDR data on scientists and engineers, including the discipline of their doctorate, their sex, and their race and ethnicity. We manually crosswalk the SDR doctorate disciplines to the MAG disciplines (see Supplementary Note 7.1 for details).

**Calculation of AI impact scores.** We estimate the direct impact of AI by implementing the "AI n-gram framework." Specifically, following prior studies[91], we identify AI-related publications using the five MAG field categories ("machine learning," "artificial intelligence," "computer vision," "natural language processing," and "pattern recognition"). Because MAG used a topic modeling approach to label each paper's field categories, the AI-related publications identified here go beyond the explicit mention of these five keywords. Identifying AI research from large-scale publication databases remains a challenging task, but the simple approach we use balances precision and recall in determining AI publications (see Supplementary Note 8.1 for details). There are also other ways to identify AI research (see Supplementary Note 2.1 for details), and our results are robust under these alternative approaches (see Supplementary Note 8 for details).

From the titles and abstracts of AI publications, we extract n-grams (bigrams and trigrams; e.g., "deep learning" and "deep neural network") and normalize them by lemmatizing and standardizing words. From these normalized n-grams, we filter AI n-grams using a list of topics under the five

AI field categories in the MAG "field of study" taxonomy. This taxonomy is constructed primarily based on Wikipedia topics (see Supplementary Note 1.1 for details). We calculate the frequency of AI n-grams per paper to approximate cumulative AI advances. Formally, the AI n-gram frequency vector at year $t$ is $\hat{G}_{AI}^t = G_{AI}^t/N_{AI}^t$, where $G_{AI}^t$ is the vector that summarizes the counts of AI n-grams extracted from AI publications before year $t$, and $N_{AI}^t$ is the number of AI publications. We repeat this process for publications in each field to extract n-grams (both AI n-grams and non-AI n-grams), and we calculate their frequency to approximate current field development. For example, the biology n-gram frequency vector at year $t$ is $\hat{G}_B^t = G_B^t/N_B^t$, where $G_B^t$ is the vector that summarizes the count of n-grams extracted from biology publications at year $t$, and $N_B^t$ is the number of these biology publications. The coordinate of the same AI n-gram in the biology frequency vector and the AI frequency vector is the same. In other words, each coordinate of $\hat{G}_B^t$ represents one n-gram, where AI n-grams have the same coordinates as those in $\hat{G}_{AI}^t$. Finally, we calculate the direct AI impact score for biology at year $t$ based on the frequency of AI n-grams:

$$S_D^t = \sum \hat{G}_B^t \cdot \hat{G}_{AI}^t, \qquad (1)$$

where the symbol "$\sum\cdot$" represents the dot product of the biology frequency vector and the AI frequency vector of the same AI n-grams. In this calculation, only their common n-grams in these two vectors are considered, and the same n-gram has the same coordinate in these two vectors. There are other ways to calculate the direct AI impact score, and our results are largely robust under some alternative calculations (see Supplementary Note 8.2 for details). A larger direct AI impact score indicates that AI has a stronger impact on the field.

We estimate the potential impact of AI by implementing the "AI capability–field task framework," which is built on the future of work literature[82-84]. It assumes that research fields may potentially benefit from AI if their basic tasks are aligned with AI capabilities (see Supplementary Note 3 for details on the underlying assumptions). We predict the capabilities of AI (i.e., what AI can do) by extracting verb-noun pairs (e.g., "learn representation") from the titles of AI publications and AI patents using a dependency parsing algorithm developed in NLP (see Supplementary Note 3.2 for details)[99-101]. Here, following the prior work[82], we only use titles because they have a higher signal-to-noise ratio than the other text fields. After normalizing verb-noun pairs through lemmatization and standardization, we calculate their relative frequency to approximate AI capabilities. Specifically, the AI capability frequency vector for AI papers at year $t$ is $Paper(\hat{C}_{AI}^t) = C_{AI}^t/\sum C_{AI}^t$, where $C_{AI}^t$ is the vector that summarizes the counts of verb-noun pairs extracted from AI publications before year $t$. We repeat this process for AI patents and calculate the vector $Patent(\hat{C}_{AI}^t)$. By taking an average of common verb-noun pairs in these two vectors, we calculate the AI capability frequency vector $\hat{C}_{AI}^t$ to approximate cumulative AI capabilities at year $t$:

$$\hat{C}_{AI}^t = [Paper(\hat{C}_{AI}^t) + Patent(\hat{C}_{AI}^t)]/2, \quad (2)$$

where the symbol "+" represents summing up the frequencies of the same verb-noun pair in the two vectors. Analogously, we predict the basic tasks of a research field (i.e., what the field does) by extracting verb-noun pairs from the titles of publications in the field. Taking the biology field as an example, the field task frequency vector at year $t$ is given by $\hat{T}_B^t = T_B^t / \sum T_B^t$, where $T_B^t$ is the vector that summarizes the counts of verb-noun pairs extracted from biology publications at year $t$. In the calculation, we apply the term frequency-inverse document frequency (tf-idf) to discount the weight of commonly appearing verb-noun pairs in both AI capability and field task vectors (see Supplementary Note 3.3 for details). Finally, we calculate the potential AI impact score of biology at year $t$ based on the alignment between its tasks and AI capabilities:

$$S_P^t = \frac{\sum \hat{T}_B^t \cdot \hat{C}_{AI}^t}{\sum \hat{C}_{AI}^t \cdot \hat{C}_{AI}^t}, \quad (3)$$

where the symbol "$\sum \cdot$" represents the dot product of the AI vector and the biology vector, and the denominator is applied to normalize the score for comparison across time. In the calculation, only common verb-noun pairs in the AI vector and the biology vector are considered, and the same verb-noun pair has the same coordinate in the two vectors. A larger potential AI impact score means that AI is predicted to have a stronger impact on the field. There are other ways to calculate the potential AI impact score, and our findings are largely robust under some alternative calculations (see Supplementary Note 8.3 for details).

**Estimation of AI education levels.** We measure the level of AI education in each discipline by leveraging OSP syllabus data and MAG publication data. This measure assumes that a discipline has a higher AI education level if a larger fraction of publications referenced by syllabi in the discipline are AI publications. The OSP dataset categorizes course syllabi by educational fields and provides a link from syllabi to their referenced publications. As syllabi with more references more likely correspond to graduate-level or research-oriented courses, we only use syllabi with at least five references and those in the recent period 2014-2018. First, we crosswalk the taxonomies of educational disciplines and academic disciplines by mapping OSP fields to MAG disciplines, and we match syllabi-referenced publications to MAG publications using the digital object identifier (DOI), title, and year (see Supplementary Note 5.1 for details). From these publications, we identify AI publications based on the MAG five AI field categories (see Supplementary Note 2.1 for details). We then estimate a discipline's AI education level by calculating the fraction of citations in the discipline's syllabi that are citations to AI publications (see Supplementary Note 5.2 for details). As robustness checks, we also use syllabi with at least ten references, calculate an alternative measure for the level of AI education defined as the fraction of a discipline's syllabi that cites at least one AI publication, and repeat the analysis for different time periods between 2000 and 2018 (see Supplementary Note 5.2 for detailed methods and additional results).

**Calculation of cross-discipline collaborations on AI.** We estimate the level of cross-discipline collaborations on AI research between domain experts and AI researchers using AI publications in each discipline other than computer science. Specifically, we first assign a primary discipline to each researcher based on the discipline in which they published most frequently in the period 1960-2019 and treat authors whose primary discipline is computer science (CS) as AI researchers (see Supplementary Note 6 for more details). We then categorize each AI publication in a discipline into one of the four co-authorship types based on its authors' primary disciplines: (1) "domain & CS," which involves both domain experts and computer scientists; (2) "domain sole," which involves only domain experts; (3) "CS sole," which involves only computer scientists; and (4) "others," which involves neither domain experts nor computer scientists. Next, we calculate the share of collaborative AI publications (i.e., those in the "domain & CS" type) for each discipline (see Supplementary Note 6.1 for detailed methods). Here the calculation only considers AI publications with at least two authors that were published in the period 1980-2019. As robustness checks, we also use an alternative approach to identify primary AI researchers (see Supplementary Note 6.3 for detailed methods and results).

## Data Availability

The MAG data are available at https://zenodo.org/record/6511057. The USPTO patent data are available at https://patentsview.org. The OSP dataset is available from the paper at https://www.pnas.org/doi/10.1073/pnas.1804247115. The SDR data are available at https://www.nsf.gov/statistics/srvydoctoratework, and the datasets used in this study are de-identified, containing only summary statistics for each discipline. The data necessary to reproduce all main plots in this paper are available at https://kellogg-cssi.github.io/ai4science.

## Code Availability

Data are linked and analyzed with customized code in Python 3 using standard software packages within these programs. The code necessary to reproduce all main plots and statistical analyses is available at https://kellogg-cssi.github.io/ai4science.

## Acknowledgements

We thank Yian Yin, Yifan Qian, Binglu Wang, Nima Dehmamy, Lav Varshney, Lili Miao, Alyse Freilich, and all members of the Center for Science of Science and Innovation (CSSI) at Northwestern University for helpful discussions. This work was supported by the Air Force Office of Scientific Research FA9550-19-1-0354, the National Science Foundation (SBE 1829344 and TIP 2241237), the Alfred P. Sloan Foundation (G-2019-12485), and the Peter G. Peterson Foundation (21048).

## Author Contributions

J.G. and D.W. conceived the idea. D.W. supervised the project. J.G. collected data and performed analyses. J.G. and D.W. analyzed the results, interpreted the findings, and wrote the paper.

## Competing Interests

The authors declare no competing interests.

# Figures

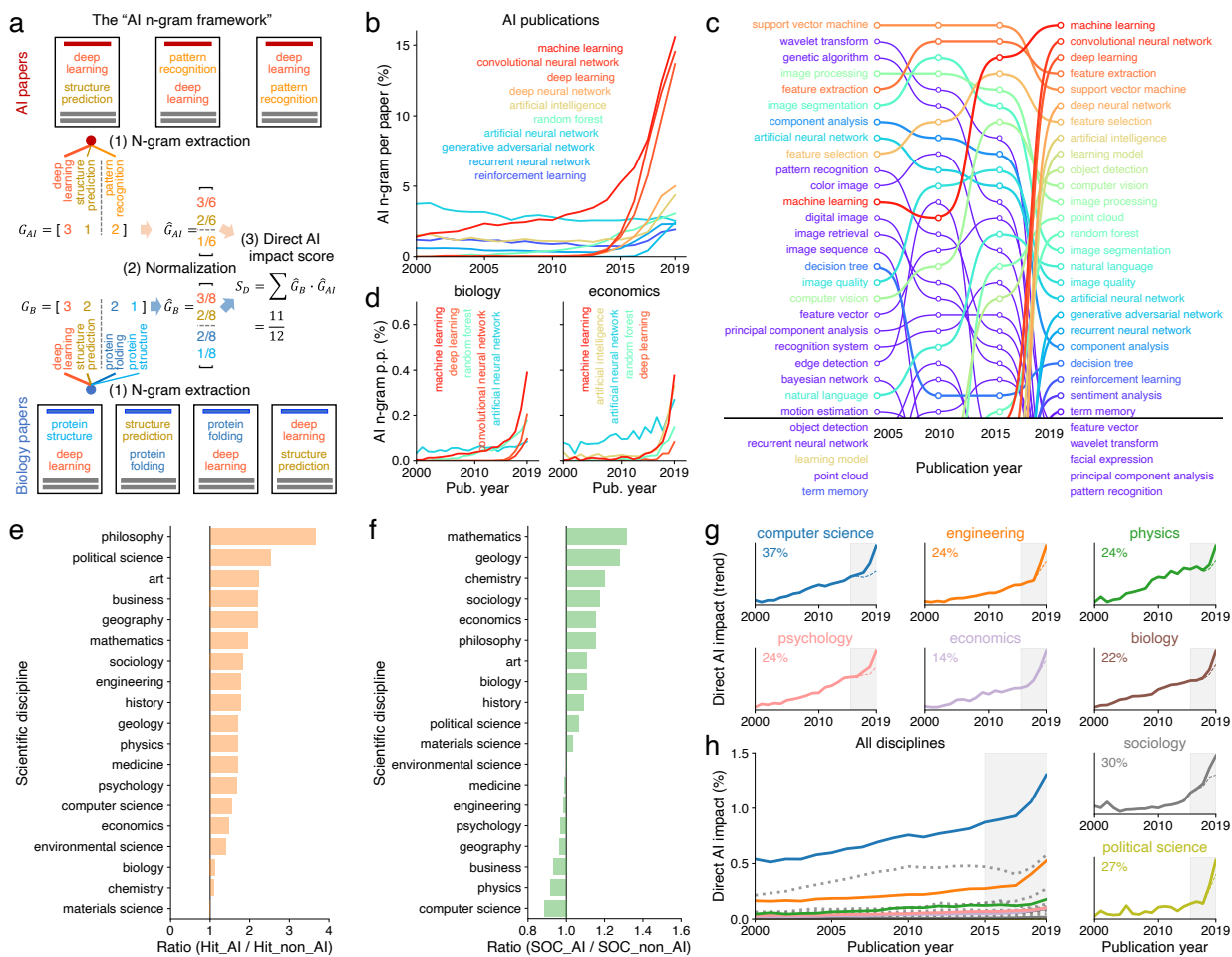

**Fig. 1. Measuring the direct impact of AI on scientific research. (a)** The "AI n-gram framework" for estimating the direct impact of AI. First, AI-related publications are identified by the MAG five AI fields. Then, n-grams are extracted from the titles and abstracts of AI publications. Next, the frequency of AI n-grams per paper is calculated after lemmatization and normalization. Similarly, n-grams are extracted for publications in each field, and the frequency of n-grams per paper is calculated. Finally, a field's direct AI impact score for a year is calculated by the dot product of the frequency of AI n-grams cumulated up to the year and the field's n-grams at the year. **(b)** The frequency of ten AI n-grams in 2019 and the trend in use of these n-grams over the past two decades. **(c)** Temporal changes in the rankings of the top 30 AI n-grams in 2019. AI n-grams are presented in rainbow color order, according to their ranking in 2019. **(d)** The frequency over the 2000-2019 time period of the top five AI n-grams in biology and economics in 2019. **(e)** The ratio of the hit rate of AI-using papers over non-AI-using papers. Here AI-using papers are identified as those that mention at least one AI n-gram, and the hit rate of papers (Hit) is defined as the likelihood that a paper is in the top 5% by total citations within the same field and year. **(f)** The ratio of the share of outside-field citations (SOC) for AI-using papers over that for non-AI-using papers. **(g)** Temporal trends in the direct AI impact scores of disciplines as shown by solid color lines. The dashed color line shows the score calculated using each discipline's yearly n-grams and AI n-grams fixed at 2015. The percentage change comparing the two scores in 2019 is shown. Each plot uses its y-axis scale to illustrate the relative change best. **(h)** The direct AI impact

scores of disciplines using the same y-axis scale. Colored lines correspond to disciplines in panel (g), and gray dotted lines represent other disciplines (see Supplementary Fig. 7 for detailed results).

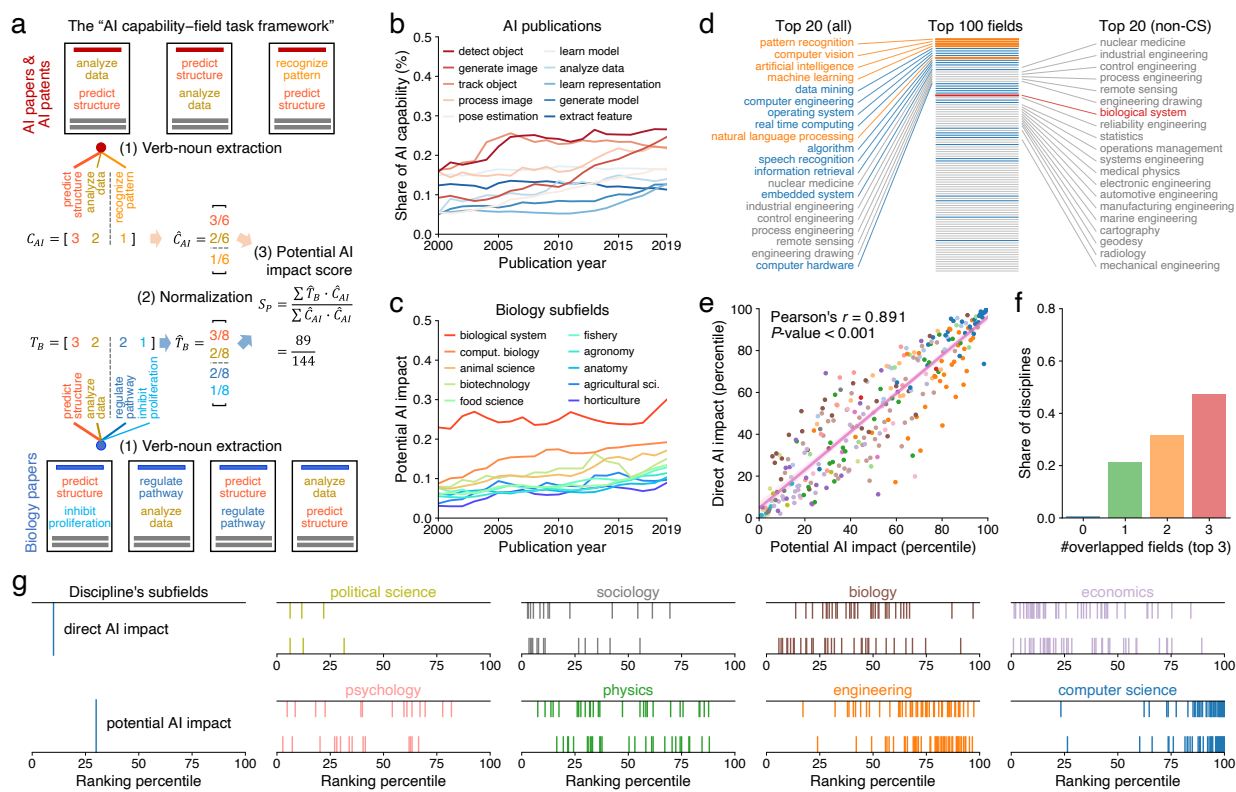

**Fig. 2. Measuring the potential impact of AI and discipline heterogeneity. (a)** The "AI capability–field task framework" for estimating the potential impact of AI. First, AI capabilities are inferred by extracting verb-noun pairs from the titles of AI publications and AI patents using a dependency parsing algorithm. Then, field tasks are inferred from publications in each field using the same method. Next, the potential AI impact score is determined by calculating the overlap between field tasks and AI capabilities, after discounting the frequency of commonly appearing verb-noun pairs. **(b)** The frequency of ten AI verb-noun pairs in 2019 and their temporal trends over the past two decades. **(c)** The top ten subfields of biology according to potential AI impact in 2019 and their temporal trend over the past two decades. The "biological system" field is consistently ranked first among all subfields. **(d)** The top 100 fields by potential AI impact in 2019 are shown with color-coded lines. The top 20 fields are listed on the left, with the five AI fields in orange, computer science (CS) fields in dark blue, and others in gray. The top 20 non-CS fields are listed on the right, with "biological system" in red and ranked seventh. **(e)** The strong correlation between the direct AI impact and potential AI impact of research fields based on their percentiles. Linear fit (centre line) with 95% confidence intervals (error bands) is shown. **(f)** A large overlap among the top three subfields for each discipline by direct AI impact and potential AI impact scores. Most disciplines exhibit three overlapped subfields (see Supplementary Table S1 for details). **(g)** The substantial heterogeneity of AI's impact within scientific disciplines. As illustrated by the legend on the left, each plot shows the percentiles of a discipline's subfields, where the percentiles based on the direct AI impact score are in the upper row and those based on the potential AI impact score are in the lower row. Ten disciplines are presented for illustration; all other disciplines are shown in Supplementary Fig. 8.

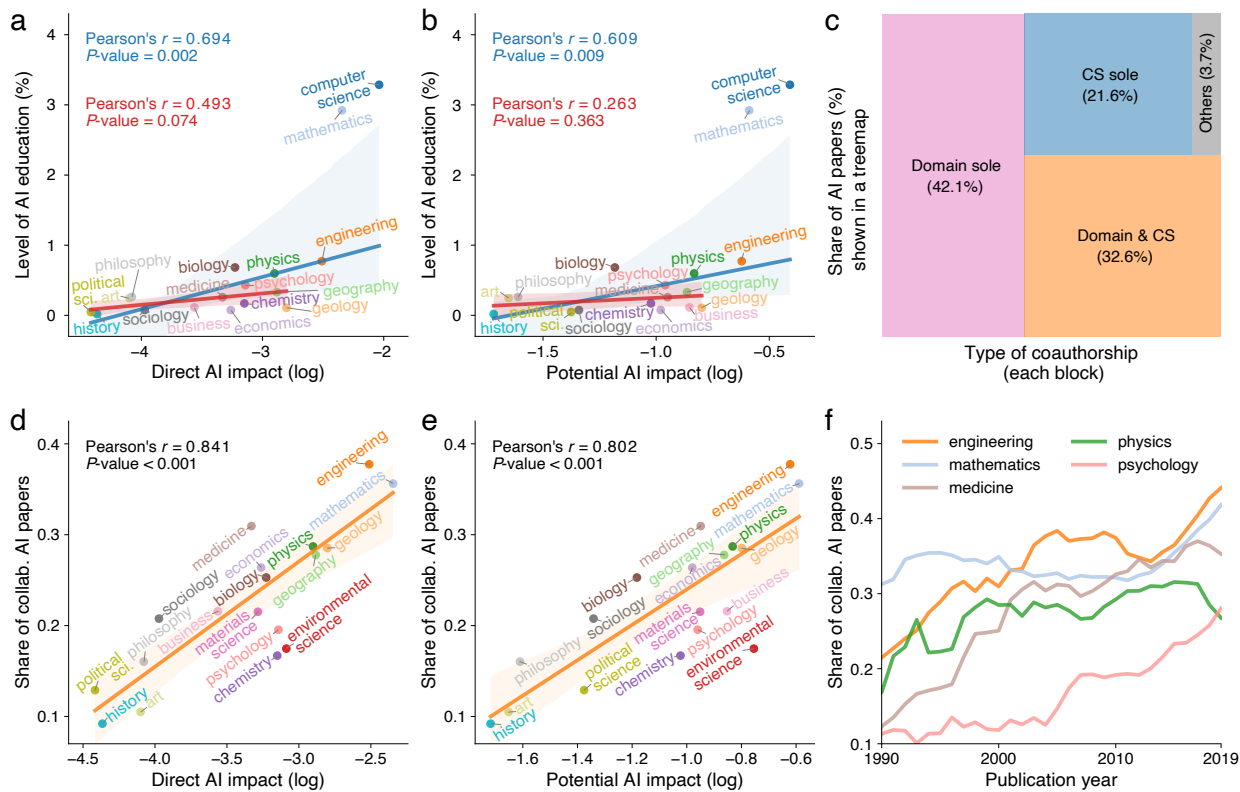

**Fig. 3. Misalignment between AI education and AI impact, but growing knowledge demand for AI. (a)** The correlation between the direct AI impact score and the AI education level that is estimated by the share of syllabus references to AI publications. Linear fits (centre lines) with 95% confidence intervals (error bands) are shown. The red line shows that the correlation loses significance when excluding the three disciplines with the largest AI impact scores: computer science, engineering, and mathematics. **(b)** The correlation between the potential AI impact score and the AI education level. Linear fits (centre lines) with 95% confidence intervals are shown. **(c)** The treemap chart shows the share of AI publications by four co-authorship types, where "domain & CS" represents collaborative AI publications by domain experts and computer scientists, "domain sole" represents AI publications by domain experts only, "CS sole" represents AI publications by computer scientists only, and "others" represents AI publications that are neither by domain experts nor by computer scientists. Here only AI publications in disciplines other than computer science with at least two authors are considered. **(d)** The positive correlation between the direct AI impact score and the share of collaborative ("domain & CS") AI publications in each discipline. Linear fit (centre line) with 95% confidence intervals (error bands) is shown. **(e)** The positive correlation between the potential AI impact score and the share of collaborative ("domain & CS") AI publications in each discipline. Linear fit (centre line) with 95% confidence intervals (error bands) is shown. **(f)** The share of collaborative AI publications ("domain & CS") in five disciplines across the period 1990-2019. Curves are smoothed by taking a three-year moving average. Results for other disciplines are shown in Supplementary Fig. 14.

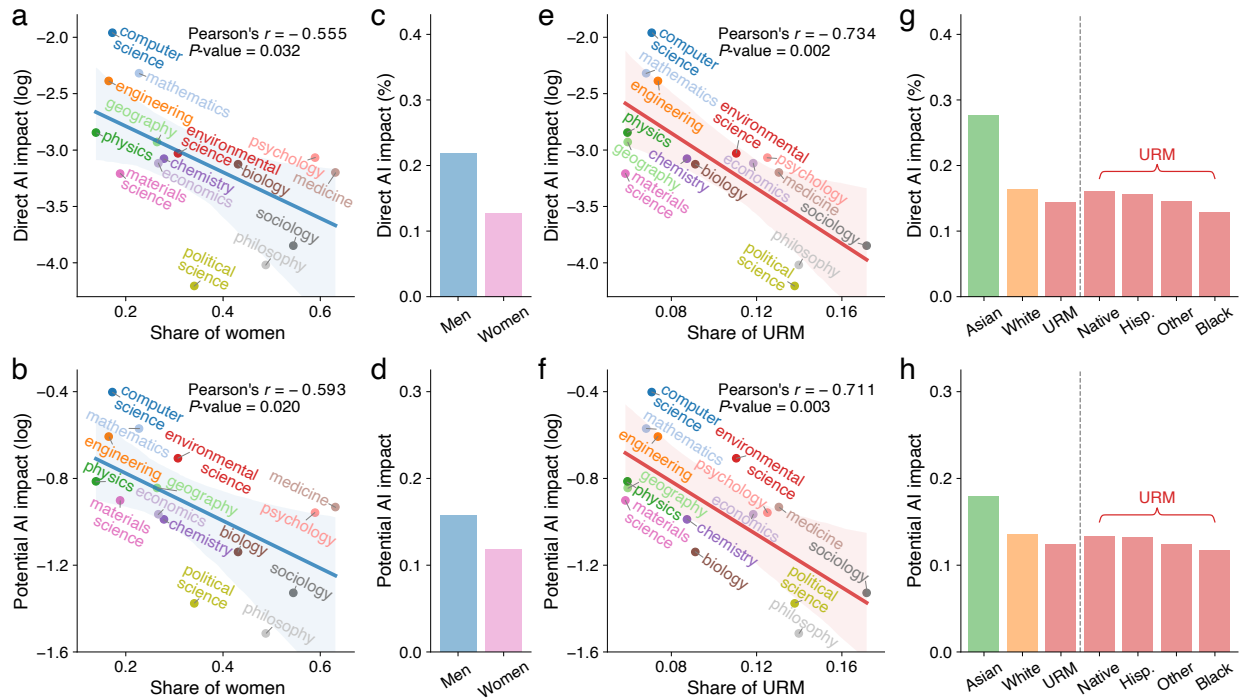

**Fig. 4. Gender and racial disparities in the impact of AI across disciplines. (a)** The negative correlation between the direct AI impact score and the share of women scientists in each discipline. Linear fit (centre line) with 95% confidence intervals (error bands) is shown. **(b)** The negative correlation between the potential AI impact score and the share of women scientists in each discipline. Linear fit (centre line) with 95% confidence intervals (error bands) is shown. **(c)** The average direct AI impact scores for women and men scientists. The average score for men/women is calculated by weighting the direct AI impact score of each discipline by the share of men/women in the discipline. **(d)** The average potential AI scores for women and men scientists. **(e)** The negative correlation between the direct AI impact score and the share of URM (underrepresented minority) scientists. The URM category includes "African American or Black," "American Indian or Alaska Native," "Hispanic or Latino," and "Native Hawaiians or other Pacific Islanders." Linear fit (centre line) with 95% confidence intervals (error bands) is shown. **(f)** The negative correlation between the potential AI impact score and the share of URM scientists. Linear fit (centre line) with 95% confidence intervals (error bands) is shown. **(g)** The average direct AI impact score for each racial and ethnic group. The average score for each group is calculated by weighting the direct AI impact score of each discipline by its share of the particular racial and ethnic group in the discipline. The average score for each racial and ethnic group under the URM category is shown separately on the right. **(h)** The average potential AI impact score for each racial and ethnic group.

# References

1   Herbert, A. S. *The Sciences of the Artificial*. (The MIT Press, 1969).
2   Brynjolfsson, E. & Mitchell, T. What can machine learning do? Workforce implications. *Science* **358**, 1530-1534 (2017).
3   Agrawal, A., Gans, J. & Goldfarb, A. *The Economics of Artificial Intelligence: An Agenda*. (University of Chicago Press, 2019).
4   Autor, D., Mindell, D. A. & Reynolds, E. B. The Work of the Future: Shaping Technology and Institutions. (MIT Task Force, 2019).
5   Acemoglu, D., Autor, D., Hazell, J. & Restrepo, P. Artificial intelligence and jobs: Evidence from online vacancies. *Journal of Labor Economics* **40**, S293-S340 (2022).
6   Aghion, P., Jones, B. F. & Jones, C. I. Artificial intelligence and economic growth. In *The Economics of Artificial Intelligence: An Agenda*   Ch. 9, 237-290 (University of Chicago Press, 2019).
7   Cockburn, I. M., Henderson, R. & Stern, S. The impact of artificial intelligence on innovation: An exploratory analysis. In *The Economics of Artificial Intelligence: An Agenda*   Ch. 4, 115-148 (University of Chicago Press, 2019).
8   Tomasev, N., Cornebise, J., Hutter, F., Mohamed, S., Picciariello, A., Connelly, B., Belgrave, D. C. M., Ezer, D., van der Haert, F. C., Mugisha, F., Abila, G., Arai, H., Almiraat, H., Proskurnia, J., Snyder, K., Otake-Matsuura, M., Othman, M., Glasmachers, T., de Wever, W., Teh, Y. W., Khan, M. E., De Winne, R., Schaul, T. & Clopath, C. AI for social good: Unlocking the opportunity for positive impact. *Nature Communications* **11**, 2468 (2020).
9   Dwivedi, Y. K., Hughes, L., Ismagilova, E., Aarts, G., Coombs, C., Crick, T., Duan, Y., Dwivedi, R., Edwards, J., Eirug, A. & others. Artificial Intelligence (AI): Multidisciplinary perspectives on emerging challenges, opportunities, and agenda for research, practice and policy. *International Journal of Information Management* **57**, 101994 (2021).
10  Frey, C. B. & Osborne, M. A. The future of employment: How susceptible are jobs to computerisation? *Technological Forecasting and Social Change* **114**, 254-280 (2017).
11  Acemoglu, D. & Restrepo, P. The race between man and machine: Implications of technology for growth, factor shares, and employment. *American Economic Review* **108**, 1488-1542 (2018).
12  Khan, H. N., Hounshell, D. A. & Fuchs, E. R. H. Science and research policy at the end of Moore's law. *Nature Electronics*. **1**, 14-21 (2018).
13  Iansiti, M. & Lakhani, K. R. *Competing in the Age of AI: Strategy and Leadership When Algorithms and Networks Run the World*. (Harvard Business Press, 2020).
14  Eshraghian, J. K. Human ownership of artificial creativity. *Nature Machine Intelligence* **2**, 157-160 (2020).
15  Marcus, G. & Davis, E. *Rebooting AI: Building Artificial Intelligence We Can Trust*. (Pantheon Books, 2019).
16  Liang, W. X., Tadesse, G. A., Ho, D., Li, F. F., Zaharia, M., Zhang, C. & Zou, J. Advances, challenges and opportunities in creating data for trustworthy AI. *Nature Machine Intelligence* **4**, 669-677 (2022).
17  Bengio, Y., Hinton, G., Yao, A., Song, D., Abbeel, P., Darrell, T., Harari, Y. N., Zhang, Y.-Q., Xue, L., Shalev-Shwartz, S., Hadfield, G., Clune, J., Maharaj, T., Hutter, F., Baydin, A. G., McIlraith, S., Gao, Q., Acharya, A., Krueger, D., Dragan, A., Torr, P., Russell, S., Kahneman, D., Brauner, J. & Mindermann, S. Managing extreme AI risks amid rapid progress. *Science* **384**, 842-845 (2024).

18  Frank, M. R., Autor, D., Bessen, J. E., Brynjolfsson, E., Cebrian, M., Deming, D. J., Feldman, M., Groh, M., Lobo, J., Moro, E., Wang, D., Youn, H. & Rahwan, I. Toward understanding the impact of artificial intelligence on labor. *Proceedings of the National Academy of Sciences, U.S.A.* **116**, 6531-6539 (2019).
19  Agrawal, A., Gans, J. S. & Goldfarb, A. Artificial intelligence: The ambiguous labor market impact of automating prediction. *Journal of Economic Perspectives* **33**, 31-50 (2019).
20  Koebis, N., Starke, C. & Rahwan, I. The promise and perils of using artificial intelligence to fight corruption. *Nature Machine Intelligence* **4**, 418-424 (2022).
21  Brynjolfsson, E., Li, D. & Raymond, L. R. Generative AI at work. *National Bureau of Economic Research*, NBER Working Paper No. w31161 (2023).
22  Noy, S. & Zhang, W. Experimental evidence on the productivity effects of generative artificial intelligence. *Science* **381**, 187-192 (2023).
23  Silver, D., Huang, A., Maddison, C. J., Guez, A., Sifre, L., van den Driessche, G., Schrittwieser, J., Antonoglou, I., Panneershelvam, V., Lanctot, M., Dieleman, S., Grewe, D., Nham, J., Kalchbrenner, N., Sutskever, I., Lillicrap, T., Leach, M., Kavukcuoglu, K., Graepel, T. & Hassabis, D. Mastering the game of Go with deep neural networks and tree search. *Nature* **529**, 484-489 (2016).
24  Geirhos, R., Temme, C. R., Rauber, J., Schütt, H. H., Bethge, M. & Wichmann, F. A. Generalisation in humans and deep neural networks. In *Advances in Neural Information Processing Systems* Vol. 31  7538-7550 (The MIT Press, 2018).
25  Grace, K., Salvatier, J., Dafoe, A., Zhang, B. & Evans, O. When will AI exceed human performance? Evidence from AI experts. *Journal of Artificial Intelligence Research* **62**, 729-754 (2018).
26  Liu, X., Faes, L., Kale, A. U., Wagner, S. K., Fu, D. J., Bruynseels, A., Mahendiran, T., Moraes, G., Shamdas, M., Kern, C. & Ledsam, J. R. A comparison of deep learning performance against health-care professionals in detecting diseases from medical imaging: A systematic review and meta-analysis. *Lancet Digital Health* **1**, e271-e297 (2019).
27  Ishowo-Oloko, F., Bonnefon, J. F., Soroye, Z., Crandall, J., Rahwan, I. & Rahwan, T. Behavioural evidence for a transparency-efficiency tradeoff in human-machine cooperation. *Nature Machine Intelligence* **1**, 517-521 (2019).
28  Yang, Y., Wu, Y. Y. & Uzzi, B. Estimating the deep replicability of scientific findings using human and artificial intelligence. *Proceedings of the National Academy of Sciences, U.S.A.* **117**, 10762-10768 (2020).
29  Wurman, P. R., Barrett, S., Kawamoto, K., MacGlashan, J., Subramanian, K., Walsh, T. J., Capobianco, R., Devlic, A., Eckert, F., Fuchs, F., Gilpin, L., Khandelwal, P., Kompella, V., Lin, H. C., MacAlpine, P., Oller, D., Seno, T., Sherstan, C., Thomure, M. D., Aghabozorgi, H., Barrett, L., Douglas, R., Whitehead, D., Durr, P., Stone, P., Spranger, M. & Kitano, H. Outracing champion Gran Turismo drivers with deep reinforcement learning. *Nature* **602**, 223-228 (2022).
30  Maslej, N., Fattorini, L., Perrault, R., Parli, V., Reuel, A., Brynjolfsson, E., Etchemendy, J., Ligett, K., Lyons, T., Manyika, J., Niebles, J. C., Shoham, Y., Wald, R. & Clark, J. The AI Index 2024 Annual Report. (AI Index Steering Committee, Institute for Human-Centered AI, Stanford University, 2024).
31  Gil, Y., Greaves, M., Hendler, J. & Hirsh, H. Amplify scientific discovery with artificial intelligence. *Science* **346**, 171-172 (2014).
32  Wang, H. C., Fu, T. F., Du, Y. Q., Gao, W. H., Huang, K. X., Liu, Z. M., Chandak, P., Liu, S. C., Van Katwyk, P., Deac, A., Anandkumar, A., Bergen, K., Gomes, C. P., Ho, S., Kohli, P., Lasenby, J., Leskovec, J., Liu, T. Y., Manrai, A., Marks, D., Ramsundar, B., Song, L., Sun,

J. M., Tang, J., Velickovic, P., Welling, M., Zhang, L. F., Coley, C. W., Bengio, Y. & Zitnik, M. Scientific discovery in the age of artificial intelligence. *Nature* **620**, 47-60 (2023).

33  Carleo, G., Cirac, I., Cranmer, K., Daudet, L., Schuld, M., Tishby, N., Vogt-Maranto, L. & Zdeborová, L. Machine learning and the physical sciences. *Reviews of Modern Physics* **91**, 045002 (2019).

34  Rahwan, I., Cebrian, M., Obradovich, N., Bongard, J., Bonnefon, J. F., Breazeal, C., Crandall, J. W., Christakis, N. A., Couzin, I. D., Jackson, M. O., Jennings, N. R., Kamar, E., Kloumann, I. M., Larochelle, H., Lazer, D., McElreath, R., Mislove, A., Parkes, D. C., Pentland, A., Roberts, M. E., Shariff, A., Tenenbaum, J. B. & Wellman, M. Machine behaviour. *Nature* **568**, 477-486 (2019).

35  Jimenez-Luna, J., Grisoni, F. & Schneider, G. Drug discovery with explainable artificial intelligence. *Nature Machine Intelligence* **2**, 573-584 (2020).

36  Xu, Y. J., Liu, X., Cao, X., Huang, C. P., Liu, E. K., Qian, S., Liu, X. C., Wu, Y. J., Dong, F. L., Qiu, C. W., Qiu, J. J., Hua, K. Q., Su, W. T., Wu, J., Xu, H. Y., Han, Y., Fu, C. G., Yin, Z. G., Liu, M., Roepman, R., Dietmann, S., Virta, M., Kengara, F., Zhang, Z., Zhang, L. F., Zhao, T. L., Dai, J., Yang, J. L., Lan, L., Luo, M., Liu, Z. F., An, T., Zhang, B., He, X., Cong, S., Liu, X. H., Zhang, W., Lewis, J. P., Tiedje, J. M., Wang, Q., An, Z. L., Wang, F., Zhang, L. B., Huang, T., Lu, C., Cai, Z. P., Wang, F. & Zhang, J. B. Artificial intelligence: A powerful paradigm for scientific research. *The Innovation* **2**, 100179 (2021).

37  Davies, A., Velickovic, P., Buesing, L., Blackwell, S., Zheng, D. I., Tomasev, N., Tanburn, R., Battaglia, P., Blundell, C., Juhasz, A., Lackenby, M., Williamson, G., Hassabis, D. & Kohli, P. Advancing mathematics by guiding human intuition with AI. *Nature* **600**, 70-74 (2021).

38  Peng, H., Ke, Q., Budak, C., Romero, D. M. & Ahn, Y. Y. Neural embeddings of scholarly periodicals reveal complex disciplinary organizations. *Science Advances* **7**, eabb9004 (2021).

39  Krenn, M., Pollice, R., Guo, S. Y., Aldeghi, M., Cervera-Lierta, A., Friederich, P., Gomes, G. D., Hase, F., Jinich, A., Nigam, A., Yao, Z. P. & Aspuru-Guzik, A. On scientific understanding with artificial intelligence. *Nature Reviews Physics.* **4**, 761-769 (2022).

40  Belikov, A. V., Rzhetsky, A. & Evans, J. A. Prediction of robust scientific facts from literature. *Nature Machine Intelligence* **4**, 445-454 (2022).

41  Grossmann, I., Feinberg, M., Parker, D. C., Christakis, N. A., Tetlock, P. E. & Cunningham, W. A. AI and the transformation of social science research. *Science* **380**, 1108-1109 (2023).

42  Groh, M., Badri, O., Daneshjou, R., Koochek, A., Harris, C., Soenksen, L. R., Doraiswamy, P. M. & Picard, R. Deep learning-aided decision support for diagnosis of skin disease across skin tones. *Nature Medicine.* **30**, 573-583 (2024).

43  Bail , C. A. Can Generative AI improve social science? *Proceedings of the National Academy of Sciences, U.S.A.* **121**, e2314021121 (2024).

44  Alvarez, A., Caliskan, A., Crockett, M. J., Ho, S. S., Messeri, L. & West, J. Science communication with generative AI. *Nature Human Behaviour.* **8**, 625-627 (2024).

45  Senior, A. W., Evans, R., Jumper, J., Kirkpatrick, J., Sifre, L., Green, T., Qin, C., Žídek, A., Nelson, A. W. R., Bridgland, A., Penedones, H., Petersen, S., Simonyan, K., Crossan, S., Kohli, P., Jones, D. T., Silver, D., Kavukcuoglu, K. & Hassabis, D. Improved protein structure prediction using potentials from deep learning. *Nature* **577**, 706-710 (2020).

46  Jumper, J., Evans, R., Pritzel, A., Green, T., Figurnov, M., Ronneberger, O., Tunyasuvunakool, K., Bates, R., Zidek, A., Potapenko, A., Bridgland, A., Meyer, C., Kohl, S. A. A., Ballard, A. J., Cowie, A., Romera-Paredes, B., Nikolov, S., Jain, R., Adler, J., Back, T., Petersen, S., Reiman, D., Clancy, E., Zielinski, M., Steinegger, M., Pacholska, M., Berghammer, T., Bodenstein, S., Silver, D., Vinyals, O., Senior, A. W., Kavukcuoglu, K.,

Kohli, P. & Hassabis, D. Highly accurate protein structure prediction with AlphaFold. *Nature* **596**, 583-589 (2021).
47  Baek, M., DiMaio, F., Anishchenko, I., Dauparas, J., Ovchinnikov, S., Lee, G. R., Wang, J., Cong, Q., Kinch, L. N., Schaeffer, R. D., Millan, C., Park, H., Adams, C., Glassman, C. R., DeGiovanni, A., Pereira, J. H., Rodrigues, A. V., van Dijk, A. A., Ebrecht, A. C., Opperman, D. J., Sagmeister, T., Buhlheller, C., Pavkov-Keller, T., Rathinaswamy, M. K., Dalwadi, U., Yip, C. K., Burke, J. E., Garcia, K. C., Grishin, N. V., Adams, P. D., Read, R. J. & Baker, D. Accurate prediction of protein structures and interactions using a three-track neural network. *Science* **373**, 871-876 (2021).
48  Popova, M., Isayev, O. & Tropsha, A. Deep reinforcement learning for de novo drug design. *Science Advances* **4**, aap7885 (2018).
49  Zhavoronkov, A., Ivanenkov, Y. A., Aliper, A., Veselov, M. S., Aladinskiy, V. A., Aladinskaya, A. V., Terentiev, V. A., Polykovskiy, D. A., Kuznetsov, M. D., Asadulaev, A. & others. Deep learning enables rapid identification of potent DDR1 kinase inhibitors. *Nature Biotechnology* **37**, 1038-1040 (2019).
50  Schneider, P., Walters, W. P., Plowright, A. T., Sieroka, N., Listgarten, J., Goodnow, R. A., Fisher, J., Jansen, J. M., Duca, J. S., Rush, T. S., Zentgraf, M., Hill, J. E., Krutoholow, E., Kohler, M., Blaney, J., Funatsu, K., Luebkemann, C. & Schneider, G. Rethinking drug design in the artificial intelligence era. *Nature Reviews Drug Discovery*. **19**, 353-364 (2020).
51  Sadybekov, A. V. & Katritch, V. Computational approaches streamlining drug discovery. *Nature* **616**, 673-685 (2023).
52  Iten, R., Metger, T., Wilming, H., Del Rio, L. d. & Renner, R. Discovering physical concepts with neural networks. *Physical Review Letters* **124**, 010508 (2020).
53  Seif, A., Hafezi, M. & Jarzynski, C. Machine learning the thermodynamic arrow of time. *Nature Physics* **17**, 105-113 (2021).
54  Wu, T. L. & Tegmark, M. Toward an artificial intelligence physicist for unsupervised learning. *Physical Review E*. **100**, 033311 (2019).
55  Lu, L., Jin, P. Z., Pang, G. F., Zhang, Z. Q. & Karniadakis, G. E. Learning nonlinear operators via DeepONet based on the universal approximation theorem of operators. *Nature Machine Intelligence* **3**, 218-229 (2021).
56  Han, J., Jentzen, A. & Weinan, E. Solving high-dimensional partial differential equations using deep learning. *Proceedings of the National Academy of Sciences, U.S.A.* **115**, 8505-8510 (2018).
57  Raayoni, G., Gottlieb, S., Manor, Y., Pisha, G., Harris, Y., Mendlovic, U., Haviv, D., Hadad, Y. & Kaminer, I. Generating conjectures on fundamental constants with the Ramanujan Machine. *Nature* **590**, 67-73 (2021).
58  Degrave, J., Felici, F., Buchli, J., Neunert, M., Tracey, B., Carpanese, F., Ewalds, T., Hafner, R., Abdolmaleki, A., de las Casas, D., Donner, C., Fritz, L., Galperti, C., Huber, A., Keeling, J., Tsimpoukelli, M., Kay, J., Merle, A., Moret, J. M., Noury, S., Pesamosca, F., Pfau, D., Sauter, O., Sommariva, C., Coda, S., Duval, B., Fasoli, A., Kohli, P., Kavukcuoglu, K., Hassabis, D. & Riedmiller, M. Magnetic control of tokamak plasmas through deep reinforcement learning. *Nature* **602**, 414-419 (2022).
59  Tshitoyan, V., Dagdelen, J., Weston, L., Dunn, A., Rong, Z., Kononova, O., Persson, K. A., Ceder, G. & Jain, A. Unsupervised word embeddings capture latent knowledge from materials science literature. *Nature* **571**, 95-98 (2019).
60  Sanchez-Lengeling, B. & Aspuru-Guzik, A. Inverse molecular design using machine learning: Generative models for matter engineering. *Science* **361**, 360-365 (2018).

61  Chen, C., Zuo, Y. X., Ye, W. K., Li, X. G., Deng, Z. & Ong, S. P. A critical review of machine learning of energy materials. *Advanced Energy Materials.* **10**, 1903242 (2020).
62  Merchant, A., Batzner, S., Schoenholz, S. S., Aykol, M., Cheon, G. & Cubuk, E. D. Scaling deep learning for materials discovery. *Nature* **624**, 80-85 (2023).
63  Zheng, S., Trott, A., Srinivasa, S., Naik, N., Gruesbeck, M., Parkes, D. C. & Socher, R. The AI Economist: Taxation policy design via two-level deep multiagent reinforcement learning. *Science Advances* **8**, eabk2607 (2022).
64  Koster, R., Jan, B., Tacchetti, A., Weinstein, A., Zhu, T. N., Hauser, O., Williams, D., Campbell-Gillingham, L., Thacker, P., Botvinick, M. & Summerfield, C. Human-centred mechanism design with Democratic AI. *Nature Human Behaviour.* **6**, 1398-1407 (2022).
65  Dunjko, V. & Briegel, H. J. Machine learning & artificial intelligence in the quantum domain: A review of recent progress. *Reports on Progress in Physics* **81**, 074001 (2018).
66  Sturm, B. L., Ben-Tal, O., Monaghan, Ú., Collins, N., Herremans, D., Chew, E., Hadjeres, G., Deruty, E. & Pachet, F. Machine learning research that matters for music creation: A case study. *Journal of New Music Research* **48**, 36-55 (2019).
67  Rajkomar, A., Dean, J. & Kohane, I. Machine learning in medicine. *New England Journal of Medicine.* **380**, 1347-1358 (2019).
68  Ramesh, A., Pavlov, M., Goh, G., Gray, S., Voss, C., Radford, A., Chen, M. & Sutskever, I. Zero-shot text-to-image generation. In *Proceedings of the 38th International Conference on Machine Learning* Vol. 139   8821-8831 (ICML, 2021).
69  Epstein, Z., Hertzmann, A. & the Investigators of Human Creativity. Art and the science of generative AI. *Science* **380**, 1110-1111 (2023).
70  Swanson, K., Liu, G., Catacutan, D. B., Arnold, A., Zou, J. & Stokes, J. M. Generative AI for designing and validating easily synthesizable and structurally novel antibiotics. *Nature Machine Intelligence* **6**, 338-353 (2024).
71  Crafts, N. Artificial intelligence as a general-purpose technology: An historical perspective. *Oxford Review of Economic Policy.* **37**, 521-536 (2021).
72  Bloom, N., Hassan, T. A., Kalyani, A., Lerner, J. & Tahoun, A. The diffusion of new technologies. *National Bureau of Economic Research*, NBER Working Paper No. w28999 (2021).
73  Caselli, F. & Coleman, W. J. Cross-country technology diffusion: The case of computers. *American Economic Review* **91**, 328-335 (2001).
74  Comin, D. & Hobijn, B. An exploration of technology diffusion. *American Economic Review* **100**, 2031-2059 (2010).
75  Zenil, H., Tegnér, J., Abrahão, F. S., Lavin, A., Kumar, V., Frey, J. G., Weller, A., Soldatova, L., Bundy, A. R., Jennings, N. R., Takahashi, K., Hunter, L., Dzeroski, S., Briggs, A., Gregory, F. D., Gomes, C. P., Rowe, J., Evans, J., Kitano, H. & King, R. The future of fundamental science led by generative closed-loop artificial intelligence. *arXiv preprint arXiv:2307.07522* (2023).
76  Topol, E. J. High-performance medicine: the convergence of human and artificial intelligence. *Nature Medicine.* **25**, 44-56 (2019).
77  Hidalgo, C. A., Orghian, D., Albo Canals, J., de Almeida, F. & Martín Cantero, N. *How Humans Judge Machines*.  (The MIT Press, 2021).
78  Raisch, S. & Krakowski, S. Artificial intelligence and management: The automation–augmentation paradox. *Academy of Management Review.* **46**, 192-210 (2021).
79  Fjelland, R. Why general artificial intelligence will not be realized. *Humanities and Social Sciences Communications* **7**, 10 (2020).
27

80 Messeri, L. & Crockett, M. J. Artificial intelligence and illusions of understanding in scientific research. *Nature* **627**, 49-58 (2024).

81 Kanitscheider, I. & Fiete, I. Training recurrent networks to generate hypotheses about how the brain solves hard navigation problems. In *Proceedings of Advances in Neural Information Processing Systems* 4529-4538 (The MIT Press, 2017).

82 Webb, M. The impact of artificial intelligence on the labor market. *Social Science Research Network*, SSRN 3482150 (2019).

83 Kogan, L., Papanikolaou, D., Schmidt, L. D. & Seegmiller, B. Technology, vintage-specific human capital, and labor displacement: Evidence from linking patents with occupations. *National Bureau of Economic Research*, NBER Working Paper No. w29552 (2022).

84 Atalay, E., Phongthiengtham, P., Sotelo, S. & Tannenbaum, D. The evolution of work in the United States. *American Economic Journal: Applied Economics* **12**, 1-34 (2020).

85 Felten, E. W., Raj, M. & Seamans, R. A method to link advances in artificial intelligence to occupational abilities. *AEA Papers and Proceedings* **108**, 54-57 (2018).

86 Wu, L., Hitt, L. & Lou, B. W. Data analytics, innovation, and firm productivity. *Management Science* **66**, 2017-2039 (2020).

87 Brynjolfsson, E., Mitchell, T. & Rock, D. What can machines learn, and what does it mean for occupations and the economy? *AEA Papers and Proceedings* **108**, 43-47 (2018).

88 Wang, D. & Barabási, A.-L. *The Science of Science*. (Cambridge University Press, 2021).

89 Fortunato, S., Bergstrom, C. T., Börner, K., Evans, J. A., Helbing, D., Milojević, S., Petersen, A. M., Radicchi, F., Sinatra, R., Uzzi, B., Vespignani, A., Waltman, L., Wang, D. & Barabási, A.-L. Science of science. *Science* **359**, eaao0185 (2018).

90 Zeng, A., Shen, Z. S., Zhou, J. L., Wu, J. S., Fan, Y., Wang, Y. G. & Stanley, H. E. The science of science: From the perspective of complex systems. *Physics Reports* **714**, 1-73 (2017).

91 Frank, M. R., Wang, D., Cebrian, M. & Rahwan, I. The evolution of citation graphs in artificial intelligence research. *Nature Machine Intelligence* **1**, 79-85 (2019).

92 Miao, L. L., Murray, D., Jung, W. S., Lariviere, V., Sugimoto, C. R. & Ahn, Y. Y. The latent structure of global scientific development. *Nature Human Behaviour.* **6**, 1206-1217 (2022).

93 Liu, L., Jones, B. F., Uzzi, B. & Wang, D. S. Data, measurement and empirical methods in the science of science. *Nature Human Behaviour.* **7**, 1046-1058 (2023).

94 Sourati, J. & Evans, J. A. Accelerating science with human-aware artificial intelligence. *Nature Human Behaviour.* **7**, 1682-1696 (2023).

95 Murray, D., Yoon, J., Kojaku, S., Costas, R., Jung, W. S., Milojevic, S. & Ahn, Y. Y. Unsupervised embedding of trajectories captures the latent structure of scientific migration. *Proceedings of the National Academy of Sciences, U.S.A.* **120**, e2305414120 (2023).

96 Krenn, M., Buffoni, L., Coutinho, B., Eppel, S., Foster, J. G., Gritsevskiy, A., Lee, H. R., Lu, Y. C., Moutinho, J. P., Sanjabi, N., Sonthalia, R., Tran, N. M., Valente, F., Xie, Y. X. Y., Yu, R. S. & Kopp, M. Forecasting the future of artificial intelligence with machine learning-based link prediction in an exponentially growing knowledge network. *Nature Machine Intelligence* **5**, 1326-1335 (2023).

97 Sinha, A., Shen, Z., Song, Y., Ma, H., Eide, D., Hsu, B. J. & Wang, K. An overview of Microsoft Academic Service (MAS) and applications. In *Proceedings of the 24th International Conference on World Wide Web* 243-246 (WWW, 2015).

98 World Intellectual Property Organization (WIPO). WIPO Technology Trends 2019 – Artificial Intelligence. (WIPO, 2019).

99 Nivre, J. & Nilsson, J. Pseudo-projective dependency parsing. In *Proceedings of the 43rd Annual Meeting of the Association for Computational Linguistics* 99-106 (ACL, 2005).

100 Honnibal, M. & Johnson, M. An improved non-monotonic transition system for dependency parsing. In *Proceedings of the 2015 Conference on Empirical Methods in Natural Language Processing* 1373-1378 (ACL, 2015).
101 Benetka, J. R., Krumm, J. & Bennett, P. N. Understanding context for tasks and activities. In *Proceedings of the 2019 Conference on Human Information Interaction and Retrieval* 133-142 (ACM, 2019).
102 Service, R. *Science's 2021 Breakthrough of the Year: Protein structures for all*, <https://www.science.org/content/article/breakthrough-2021> (2021).
103 Börner, K., Scrivner, O., Gallant, M., Ma, S., Liu, X., Chewning, K., Wu, L. & Evans, J. A. Skill discrepancies between research, education, and jobs reveal the critical need to supply soft skills for the data economy. *Proceedings of the National Academy of Sciences, U.S.A.* **115**, 12630-12637 (2018).
104 Wuchty, S., Jones, B. F. & Uzzi, B. The increasing dominance of teams in production of knowledge. *Science* **316**, 1036-1039 (2007).
105 Wu, L. F., Wang, D. S. & Evans, J. A. Large teams develop and small teams disrupt science and technology. *Nature* **566**, 378-382 (2019).
106 Littmann, M., Selig, K., Cohen-Lavi, L., Frank, Y., Honigschmid, P., Kataka, E., Mosch, A., Qian, K., Ron, A., Schmid, S., Sorbie, A., Szlak, L., Dagan-Wiener, A., Ben-Tal, N., Niv, M. Y., Razansky, D., Schuller, B. W., Ankerst, D., Hertz, T. & Rost, B. Validity of machine learning in biology and medicine increased through collaborations across fields of expertise. *Nature Machine Intelligence* **2**, 18-24 (2020).
107 Mehrabi, N., Morstatter, F., Saxena, N., Lerman, K. & Galstyan, A. A survey on bias and fairness in machine learning. *ACM Computing Surveys.* **54**, 1-35 (2021).
108 Young, E., Wajcman, J. & Sprejer, L. Where are the Women? Mapping the gender job gap in AI. (The Alan Turing Institute, 2021).
109 Xie, Y. & Shauman, K. A. *Women in science: Career processes and outcomes*. (Harvard University Press, 2003).
110 Hoppe, T. A., Litovitz, A., Willis, K. A., Meseroll, R. A., Perkins, M. J., Hutchins, B. I., Davis, A. F., Lauer, M. S., Valantine, H. A., Anderson, J. M. & Santangelo, G. M. Topic choice contributes to the lower rate of NIH awards to African-American/black scientists. *Science Advances* **5**, eaaw7238 (2019).
111 Ginther, D. K., Schaffer, W. T., Schnell, J., Masimore, B., Liu, F., Haak, L. L. & Kington, R. Race, ethnicity, and NIH research awards. *Science* **333**, 1015-1019 (2011).
112 Sugimoto, C. R., Larivière, V., Ni, C. Q., Gingras, Y. & Cronin, B. Global gender disparities in science. *Nature* **504**, 211-213 (2013).
113 Huang, J. M., Gates, A. J., Sinatra, R. & Barabási, A. L. Historical comparison of gender inequality in scientific careers across countries and disciplines. *Proceedings of the National Academy of Sciences, U.S.A.* **117**, 4609-4616 (2020).
114 The National Network for Critical Technology Assessment (NNCTA). Securing America's Future: A Framework for Critical Technology Assessment. (NNCTA, 2023).
115 Cachola, I., Lo, K., Cohan, A. & Weld, D. S. TLDR: Extreme summarization of scientific documents. In *Proceedings of the 2020 Conference on Empirical Methods in Natural Language Processing* 4766-4777 (ACL, 2020).
116 Lew, A., Agrawal, M., Sontag, D. & Mansinghka, V. PClean: Bayesian data cleaning at scale with domain-specific probabilistic programming. In *Proceedings of the 24th International Conference on Artificial Intelligence and Statistics* Vol. 130 1927-1935 (JMLR, 2021).

117 Bommasani, R., Hudson, D. A., Adeli, E., Altman, R., Arora, S., von Arx, S., Bernstein, M. S., Bohg, J., Bosselut, A., Brunskill, E. & Brynjolfsson, E. On the opportunities and risks of foundation models. *arXiv preprint arXiv:2108.07258* (2021).

118 Wei, J., Tay, Y., Bommasani, R., Raffel, C., Zoph, B., Borgeaud, S., Yogatama, D., Bosma, M., Zhou, D., Metzler, D., Chi, E. H., Hashimoto, T., Vinyals, O., Liang, P., Dean, J. & Fedus, W. Emergent abilities of large language models. *arXiv preprint arXiv:2206.07682* (2022).

119 Moor, M., Banerjee, O., Abad, Z. S. H., Krumholz, H. M., Leskovec, J., Topol, E. J. & Rajpurkar, P. Foundation models for generalist medical artificial intelligence. *Nature* **616**, 259-265 (2023).

120 Huang, Z., Bianchi, F., Yuksekgonul, M., Montine, T. J. & Zou, J. A visual-language foundation model for pathology image analysis using medical Twitter. *Nature Medicine.* **29**, 2307-2316 (2023).

121 Goldfarb, A., Taska, B. & Teodoridis, F. Could machine learning be a general purpose technology? A comparison of emerging technologies using data from online job postings. *Research Policy* **52**, 104653 (2023).

122 Fawzi, A., Balog, M., Huang, A., Hubert, T., Romera-Paredes, B., Barekatain, M., Novikov, A., Ruiz, F. J. R., Schrittwieser, J., Swirszcz, G., Silver, D., Hassabis, D. & Kohli, P. Discovering faster matrix multiplication algorithms with reinforcement learning. *Nature* **610**, 47-53 (2022).

123 Jobin, A., Ienca, M. & Vayena, E. The global landscape of AI ethics guidelines. *Nature Machine Intelligence* **1**, 389-399 (2019).

124 Arrieta, A. B., Díaz-Rodríguez, N., Del Ser, J., Bennetot, A., Tabik, S., Barbado, A., García, S., Gil-López, S., Molina, D., Benjamins, R., Chatila, R. & Herrera, F. Explainable Artificial Intelligence (XAI): Concepts, taxonomies, opportunities and challenges toward responsible AI. *Information Fusion.* **58**, 82-115 (2020).

125 Lenharo, M. An AI revolution is brewing in medicine. What will it look like? *Nature* **622**, 686-688 (2023).

126 Bockting, C. L., van Dis, E. A. M., van Rooij, R., Zuidema, W. & Bollen, J. Living guidelines for generative AI—why scientists must oversee its use. *Nature* **622**, 693-696 (2023).

127 Schwartz, I. S., Link, K. E., Daneshjou, R. & Cortes-Penfield, N. Black box warning: Large language models and the future of infectious diseases consultation. *Clinical Infectious Diseases.* **78**, 860-866 (2024).

128 Ahmadpoor, M. & Jones, B. F. The dual frontier: Patented inventions and prior scientific advance. *Science* **357**, 583-587 (2017).

129 Mukherjee, S., Romero, D. M., Jones, B. & Uzzi, B. The nearly universal link between the age of past knowledge and tomorrow's breakthroughs in science and technology: The hotspot. *Science Advances* **3**, e1601315 (2017).

130 Yin, Y. A., Dong, Y. X., Wang, K. S., Wang, D. S. & Jones, B. F. Public use and public funding of science. *Nature Human Behaviour.* **6**, 1344-1350 (2022).

# Supplementary Information:

# Quantifying the Benefit of Artificial Intelligence for Scientific Research

Jian Gao[1,2,3], Dashun Wang[1,2,3,4*]

[1] Center for Science of Science and Innovation, Northwestern University, Evanston, IL, 60208, USA
[2] Kellogg School of Management, Northwestern University, Evanston, IL, 60208, USA
[3] Northwestern Institute on Complex Systems, Northwestern University, Evanston, IL, 60208, USA
[4] McCormick School of Engineering, Northwestern University, Evanston, IL, 60208, USA
* Correspondence to: dashun.wang@northwestern.edu

## Table of Contents

# Supplementary Note 1. Data description

## 1.1 Microsoft Academic Graph (MAG) publication data

The publication data are sourced from the Microsoft Academic Graph (MAG) database [1-3], which is among the largest open-source publication databases in the world. The MAG dataset contains records of about 200 million documents of various types (e.g., papers and books), interlinked through citation relationships. For each document, the data include information about the author, affiliation, publication venue, and research topics. We use the December 2019 version of the MAG dataset to avoid any overlap with the period of the COVID-19 pandemic, which may have had a substantial impact on the publication and collaboration landscape, especially in research that is not related to COVID-19 [4-7]. While the MAG database has been retired as of December 2021, its replacement and successor OpenAlex draws data from MAG's existing records and other sources, providing a free, up-to-date version [8, 9]. In addition, there is a publicly available, large-scale data lake called SciSciNet, which pre-processes the last MAG data snapshot (from December 2021) and provides millions of external linkages to other data sources [10]. At the time of this writing, a version of the MAG data (from September 2021) is also freely accessible at Zenodo [11].

To focus on scientific publications, we only consider MAG documents that are categorized as "journal," "conference," "book," or "book chapter." All other documents are excluded from the analysis, including those categorized as "patent" and "thesis." We collect information on the author, title, abstract, publication year, field of study, and digital object identifier (DOI) for each publication, as well as the citation relationships among these publications. The MAG "field of study" taxonomy contains six levels (from L0 to L5), covering more than 700K concepts, which are not only seeded from Wikipedia pages (in which each page introduces a concept) and the Unified Medical Language System (UMLS) vocabulary but also automatically discovered from the MAG literature using some embedding models [12, 13]. Under the "field of study" taxonomy, publications are categorized into 19 broad disciplines at the L0 level (e.g., "computer science") and 292 granular fields at the L1 level (e.g., "machine learning"). Each publication is often assigned to one L0-level scientific discipline and several L1-level research fields with different scores. For those belonging to multiple disciplines or fields, we consider the one with the highest score at each level. Further, we filter documents published during the period from 1960 to 2019, those with English titles and abstracts, and those that have a "field of study" at both the L0 level and L1 level. When there are multiple versions of the same paper, we only keep the primary record that is identified by MAG. These data cleaning and filtering processes lead to a subset of 74.6 million publications in the analysis.

**1.2 United States Patent and Trademark Office (USPTO) patent data**

The patent data are collected from PatentsView [14, 15], a data platform based on bulk data from the United States Patent and Trademark Office (USPTO) [16]. We use the December 2019 version of the PatentsView USPTO patent data. For each patent document, the data provide information about the title, abstract, year granted, technology classes, inventor, and details of the invention. Each patent is associated with a list of technology classification codes that are based on the Cooperative Patent Classification (CPC) system. We use these CPC codes, supplemented with a list of AI-related keywords, to identify AI patents (see Supplementary Note 3.1 for details). We filter patents that were granted in the period from 1976 to 2019 and those with titles and CPC codes. This data filtering process leads to a subset of 7.1 million patents in the analysis.

**1.3 Open Syllabus Project (OSP) syllabus data**

The syllabus data are sourced from the Open Syllabus Project (OSP) [17-19], which is the world's first large-scale online database of university course syllabi, covering about nine million English-language syllabi from 140 countries. Each syllabus document contains the summary text, year, field name, Classification of Instructional Programs (CIP) code, and a list of scientific references. As our focus is on recent educational trends, we filter syllabus documents published in the period from 2000 to 2018, leading to a subset of 4.2 million course syllabi. Both field names and CIP codes in a syllabus document indicate the educational discipline of the course. Therefore, we crosswalk these two types of educational disciplines to the research disciplines in MAG (see Supplementary Note 5.1 for details). We rely on field names in the analysis and use CIP codes as robustness checks. Each scientific reference contains information about the author, title, journal, year, and DOI. Most references cited in syllabus documents are research papers and books, which we match with MAG publication records (see Supplementary Note 5.1 for details).

**1.4 Survey of Doctorate Recipients (SDR) demographic data**

The demographic data on scientists and researchers are collected from the Survey of Doctorate Recipients (SDR) [20], which provides demographic information for individuals with a U.S.

research doctoral degree in a science, engineering, or health field. We rely on the SDR data for three primary reasons. First, the data provide reliable survey-based demographic statistics on the sex, ethnicity, and race of researchers. By comparison, name-based gender and race inference algorithms often fail to achieve a desirable level of accuracy, as has been shown, for instance, for author names with East Asian origins [21-23]. Second, the data cover a substantial portion of the international research workforce. Over 10% of U.S.-trained science, engineering, and health doctorate holders (about 1.1 million individuals) were asked to participate in the 2017 SDR, and they live and work throughout the world [24]. Third, the SDR provides statistical data by the field of doctorate, which is the granularity required to be manually cross-walked to MAG scientific disciplines (see Supplementary Note 7.1 for details). Our study, therefore, uses summary statistics from the 2017 SDR data, which were de-identified, covering 815K U.S.-residing employed doctoral scientists and engineers, by the field of doctorate, sex, and ethnicity/race.

## Supplementary Note 2. Measuring direct AI impact

### 2.1 Identification of AI-related publications and n-grams

Identifying AI-related research from large-scale publication databases is a challenging task, as the definitions of AI are often ambiguous and vary by context [25-33]. Following prior works [34-36], we identify AI-related publications from the MAG datasets using the five L1-level field categories: "machine learning," "artificial intelligence," "computer vision," "natural language processing," and "pattern recognition." These five benchmarking AI fields all belong to the L0-level "computer science" discipline according to the MAG "field of study" formal hierarchy [12]. Publications within any of these five AI fields are seen as AI-related papers in the analysis. Because MAG used a topic modeling approach to label the field for each paper, the AI papers we identify using the five field categories go beyond the explicit mention of these keywords [1-3]. We find that the share of AI-related papers among all papers published in each discipline has increased for most disciplines, especially in recent years (Supplementary Figure 1).

In addition, we identify a broader set of papers that mention AI-related terms. Specifically, using Natural Language Processing (NLP) techniques [37-39], we extract n-grams (bigrams and trigrams) from the titles and abstracts of AI-related publications that are initially identified by the five MAG benchmarking AI fields, and we normalize them by lemmatizing and standardizing words. From these n-grams, we further filter AI n-grams using a list of concepts that belong to the five AI fields according to the MAG "field of study" formal hierarchy [12]. This process helps filter out n-grams that are not meaningful for scientific research. Some of the remaining n-grams are still very generic, so we further hand-curated a list of AI-related n-grams. We apply this n-gram approach to all publications and identify those that mention at least one AI n-gram as publications that use AI. This approach allows us to identify AI-related research not only in the five benchmarking AI fields and the "computer science" discipline but also in other fields and disciplines.

There are several limitations and alternatives to this identification approach. Arguably, papers that belong to the five benchmarking AI fields in the MAG dataset or mention AI terms in their titles or abstracts do not necessarily use AI or benefit from AI advances; it is possible that some of these papers simply discuss AI. It would ideally to scan the full text of the publications with emphasis on the method sections. However, systematic data on the full text or the method sections are not currently available. While our approach may, therefore, miss some AI-related papers because we only use the title and abstract of each publication, the criteria we use serve as an imperfect proxy for AI research. We perform a precision-recall analysis of the approach, finding that it has high precision and recall in identifying AI-related research (see Supplementary Note 8.1 for details). We also use two alternative approaches to identify AI research: (1) using a broader set of AI keywords as MAG field categories and (2) using a set of four arXiv subject categories. We find that our main results are largely robust (see Supplementary Note 8.1 for details).

## 2.2 Temporal trends in AI n-grams in disciplines

We apply the n-gram approach to AI papers, identify a list of AI n-grams, and then apply the approach to all MAG publications in each discipline and field. We find that the landscape of AI research is evolving dynamically. Specifically, some conventional AI n-grams have become less common over time, some were dormant for a long period but rose to prominence in recent years, and some novel AI n-grams emerged only recently (see Figures 1b,c in the main text). At the same time, we find that AI n-grams are increasingly mentioned in papers in disciplines and fields outside of AI over time, suggesting that AI is having a progressively broader impact across the sciences. The increasing trend of AI n-grams in publications from different disciplines is especially evident in recent years, primarily after 2015. This pattern is observed not only in the Science, Technology, Engineering, and Math (STEM) or computational disciplines such as physics, medicine, biology, and chemistry, but also in some social science disciplines such as business, psychology, and sociology (Supplementary Figure 2).

## 2.3 Calculation of the direct AI impact score

The dynamic landscape of AI research prompts us to study the use of AI in different fields and the direct impact of AI (i.e., the relative frequency of AI methods uptake in science) over time. To this end, we estimate the direct impact of AI by implementing the "AI n-gram framework." Specifically, we first extract AI n-grams from the titles and abstracts of AI publications and calculate the frequency of AI n-grams per paper to approximate cumulative AI advances in research. Formally, the frequency vector of AI n-grams at year $t$ is given by

$$\hat{G}_{AI}^t = \frac{G_{AI}^t}{N_{AI}^t},$$

where $G_{AI}^t$ is the vector that summarizes the counts of AI n-grams extracted from AI papers published before year $t$, and $N_{AI}^t$ is the number of AI papers. Here we use 1960 as the starting year to cumulate AI n-grams. For papers in each field, we apply the same approach to extract n-grams (AI n-grams and non-AI n-grams) and calculate their frequency to approximate field development. Formally, the n-gram frequency vector for the biology discipline at year $t$ is given by

$$\hat{G}_B^t = \frac{G_B^t}{N_B^t},$$

where $G_B^t$ is the vector that summarizes the count of n-grams extracted from biology papers published at year $t$, and $N_B^t$ is the number of these biology papers. The coordinate of the same AI n-gram in the biology frequency vector and the AI frequency vector is the same. In other words, each coordinate of the normalized biology vector $\hat{G}_B^t$ represents one n-gram, where AI n-grams have the same coordinates as those in the normalized AI vector $\hat{G}_{AI}^t$. Finally, we calculate the direct AI impact score for biology at year $t$ using a dot product measure:

$$S_D^t = \sum \hat{G}_B^t \cdot \hat{G}_{AI}^t,$$

where the symbol "$\sum \cdot$" represents the dot product among the biology frequency vector and the AI frequency vector of the same AI n-grams. In this calculation, only common n-grams in the AI n-gram vector and the biology n-gram vector are considered, and the same n-gram has the same coordinate in these two vectors. A larger direct AI impact score indicates that more AI n-grams are used in the research field and, thus, that it receives a larger direct benefit from AI.

We apply the "AI n-gram framework" to the titles and abstracts of publications between 1960 and 2019. There, we calculate the direct AI impact score for each of the 19 MAG L0-level scientific disciplines (e.g., "biology") and year. Applying the same approach, we calculate the direct AI impact score for each of the 292 MAG L1-level research fields (e.g., "biological system") and year. At the scientific discipline level, we find that the direct AI impact score increased over time in the period 1960-2019, and there has been a notably sharp increase since 2015 across many disciplines, suggesting the widespread impact of AI on scientific research (Supplementary Figure 3). In addition to the dot product, we tried two alternative measures to calculate the direct AI impact score, finding that our main results are largely robust (see Supplementary Note 8.2 for details).

## 2.4 Citation premium of papers that mention AI terms

Based on the curated data on AI n-grams, we explore the extent to which using AI is associated with a citation premium. Specifically, we calculate the average rate for AI papers (i.e., those that mention at least one AI n-gram) to be "hit papers," which we define as being in the top 5% by total citations in the same field and year. We find that AI papers are more likely to become hit papers than non-AI papers in each discipline (Supplementary Figure 4a). For a majority of fields, the hit rate of AI papers is larger than the baseline 5%, and it has a significantly negative correlation (two-sided Pearson's correlation test; Pearson's $r = -0.378$; $P$-value < 0.001) with the direct AI impact score (Supplementary Figure 4b). These results suggest that AI papers in research fields with less direct AI impact tend to be associated with a larger citation premium on average.

Moreover, we study the broader citation impact of papers that use AI by calculating the share of citations AI papers receive that are outside-field citations. At the discipline level, we find that AI papers have a larger share of outside-field citations than non-AI papers (Supplementary Figure 4c). At the same time, there is substantial heterogeneity across disciplines, as AI papers receive disproportionately more outside-field citations in biology, geology, and arts than non-AI papers. At the field level, we find that the ratio of AI to non-AI papers for the share of outside-field citations is negatively correlated (two-sided Pearson's correlation test; Pearson's $r = -0.215$; $P$-value < 0.001) with the direct AI impact score (Supplementary Figure 4d). Taken together, these results suggest that papers that use AI tend to have a larger and broader citation impact on science than traditional field papers. Moreover, the citation premium of papers that use AI appears to be stronger for disciplines with an overall lower propensity to use AI, suggesting that disciplines that seem distant from AI may reap substantial benefits from using AI to advance their research work. It should be noted that the analysis does not claim that citations result from the use of AI in the research. Instead, it highlights the citation premium of papers that mention AI terms in their titles or abstracts, which is a correlation and does not imply causality.

## Supplementary Note 3. Measuring potential AI impact

### 3.1 Identification of AI-related patents

The identification of AI-related patents, like the identification of AI-related papers, is a challenging task [29, 40-43]. Among the USPTO patents collected from PatentsView, we identify AI-related patents using a mixed method of keywords and patent CPC codes, as suggested by the World Intellectual Property Organization (WIPO) [44]. The AI-related keywords (key phrases) and CPC codes are sourced from the PATENTSCOPE Artificial Intelligence Index [45]. The index provides AI techniques, AI functional applications, and AI application fields, which have been used to carry out searches in PATENTSCOPE and other patent databases to identify patent applications related to AI. Following this line of work, we identify AI-related patents in the USPTO data by matching CPC codes and searching for key AI phrases in the titles of patents. Altogether, we identify 132K AI-related patents that were granted between 1976 and 2019. There are, however, other ways to identify AI-related patents, including the Artificial Intelligence Patent Dataset (AIPD) that results from a machine learning approach [41] and the AI technological innovation dataset that results from NLP and machine learning tools for text classification [46], which became available recently.

### 3.2 Inference of AI capabilities and field tasks

While the explicit mention of AI-related n-grams signals direct AI benefits, it is possible that research fields could be using AI to perform more of the basic tasks in their scientific research, suggesting the potential impact of AI on science. To this end, building on the future of work literature [47-54], we develop the "AI capability–field task framework" to estimate the potential AI impact on scientific research. This framework relies on the inference of AI capabilities from AI publications and patents and field tasks from publications.

We infer the capabilities of AI (i.e., the tasks that AI can perform) and the basic tasks of research fields (i.e., the tasks undertaken by a research field) by applying NLP techniques to analyze publications and patents. Specifically, we first employ a dependency parsing algorithm (spaCy dependency parser in the Python library [38, 55]) to extract verb-noun pairs from the titles of AI publications that are identified by the five MAG AI fields. We then lemmatize verbs and nouns in these verb-noun pairs, and we group words with similar meanings into one word that represents the meaning. For example, verbs like "detecting," "detected," and "detects" are represented by "detect." This cleaning and normalization process through lemmatization and standardization reduces noise in identifying relevant verb-noun pairs from publications. As an illustrative example, from the AI paper titled, "Using Bayesian networks to analyze medical data" [56], the algorithm extracts "use network" and "analyze data" as the verb-noun pairs (Supplementary Figure 5).

We apply the verb-noun extraction method to AI publications identified by the five MAG AI fields and AI patents identified by keywords and CPC codes. The verb-noun pairs extracted from AI publications and patents together represent the capabilities of AI (see the word cloud of AI verb-noun pairs cumulative up to 2019 in Supplementary Figure 6a). We apply the same approach to analyze publications in each field. The verb-noun pairs extracted from a field's publications represent the basic tasks of the field (see the word clouds of verb-noun pairs in the biology and chemistry disciplines in 2019 in Supplementary Figure 6b and Supplementary Figure 6c, respectively). As some of the verb-noun pairs extracted from publications and patents are very generic, we inspected and manually curated a list of verbs, nouns, and verb-noun pairs for frequent words. It is important to note that here we only extract verb-noun pairs from the titles of publications and patents because they have a much higher signal-to-noise ratio than the other publication text fields, such as the abstract, and the other patent text fields, such as the abstract, description, and claims [47].

### 3.3 Calculation of potential AI impact score

The "AI capability–field task framework" assumes that research fields may potentially benefit from AI if their basic tasks are aligned with AI capabilities. To this end, we estimate the potential AI impact score based on the alignment between AI capabilities and field tasks. Specifically, we first calculate the relative frequency of verb-noun pairs extracted from the titles of AI papers and AI patents to approximate cumulative AI capabilities for each. Formally, the AI capability frequency vector for papers at year $t$ is given by

$$Paper(\hat{C}_{AI}^t) = \frac{C_{AI}^t}{\sum C_{AI}^t},$$

where $C_{AI}^t$ is the vector that summarizes the counts of verb-noun pairs extracted from AI papers before year $t$. We apply the term frequency-inverse document frequency (tf-idf) to discount the weight of commonly appearing pairs [57]. Here, each vector coordinate is one verb-noun pair, and the value is its relative frequency. Likewise, we repeat this process for AI patents and calculate the AI capability frequency vector $Patent(\hat{C}_{AI}^t)$. We further normalize frequency vectors of AI papers and AI patents to approximate cumulative AI capabilities at year $t$. Specifically, we take an average of common verb-noun pairs in these two frequency vectors, where the same verb-noun pair shares the same coordinate. Formally, the AI capability frequency vector at year $t$ is given by:

$$\hat{C}_{AI}^t = \frac{[Paper(\hat{C}_{AI}^t) + Patent(\hat{C}_{AI}^t)]}{2},$$

where the symbol "+" represents summing two frequencies of the same verb-noun pair. For simplicity and following prior works [58], here we treat verb-noun pairs extracted from AI papers and AI patents as equally important in inferring AI capabilities, and we normalize them separately. In the calculation, we use 1960 as the starting year to cumulate AI capabilities based on AI publications and AI patents. Because there is no patent data in the period 1960-1975, we only use $Patent(\hat{C}_{AI}^t)$ to represent the AI capability frequency vector $\hat{C}_{AI}^t$ for this period.

We then calculate the relative frequency of verb-noun pairs extracted from the titles of papers in a research field to approximate the current basic tasks of the field. Formally, taking the biology discipline (or one of its subfields) as an example, the field task frequency vector at year $t$ is

$$\hat{T}_B^t = \frac{T_B^t}{\sum T_B^t},$$

where $T_B^t$ is the vector that summarizes the counts of verb-noun pairs extracted from biology papers published at year $t$. We also apply the tf-idf approach to discount the weight of commonly appearing pairs in field tasks. Finally, we estimate the potential AI impact score of biology at year $t$ by calculating the overlap between its field tasks and cumulative AI capabilities at $t$:

$$S_P^t = \frac{\sum \hat{T}_B^t \cdot \hat{C}_{AI}^t}{\sum \hat{C}_{AI}^t \cdot \hat{C}_{AI}^t},$$

where the symbol "$\sum \cdot$" represents the dot product among the AI verb-noun vector and the biology verb-noun vector. Specifically, only common verb-noun pairs in the AI verb-noun vector and the biology verb-noun vector are considered in the calculation, and the same verb-noun pair has the same coordinate in these two vectors. The denominator is applied to normalize the AI score for comparison across time. The normalization assumes that AI can perform all research tasks that are inferred from AI publications in the five AI fields.

A larger potential AI impact score for a field indicates a better alignment between field tasks and AI capabilities and thus a higher potential for the field to benefit from AI advances. Applying the "AI capability–field task framework" to all publications between 1960 and 2019, we calculate the potential AI impact score for each of the 19 disciplines and years. Applying the same approach, we calculate the potential AI impact score for each of the 292 fields and years. In addition to the dot product, we tried two alternative measures to calculate the potential AI impact score, finding that our main results are largely robust (see Supplementary Note 8.3 for details).

# Supplementary Note 4. Unpacking AI's growing impact

## 4.1 Discipline heterogeneity and field consistency

We estimate the direct AI impact using the "AI n-gram framework" and the potential AI impact using the "AI capability–field task framework" based on publications and patents. At the discipline level, we find that the differences in the direct AI impact scores span as large as two magnitudes (Supplementary Figure 7a), suggesting that there is substantial heterogeneity across disciplines. By comparison, the magnitude of discipline-level differences for the potential AI impact scores is smaller, despite the rapid increase in the potential AI impact scores of some disciplines, such as material science and chemistry (Supplementary Figure 7b).

We further explore the within-discipline heterogeneity by plotting the positions of a discipline's child fields that are ranked in percentile by their direct and potential AI scores (Supplementary Figure 8). In the percentiles of 292 fields, the field with the lowest AI score is $0^{th}$, and the field with the highest AI score is $100^{th}$. We find that the percentiles of the direct AI impact score and potential AI impact score for all fields are highly correlated with each other (see Fig. 2e in the main text), and the two percentiles for the same subfield within each discipline are largely consistent (Supplementary Figure 8). Moreover, we find that almost every discipline contains some subfields that are greatly impacted by AI—even those that have relatively small AI impact scores. For example, "medicine" as a discipline has medium AI impact scores, but it contains some subfields that have large AI impact scores, such as "nuclear medicine," "optometry," and "medical physics" (Supplementary Table 1). These results suggest that the benefits of AI for scientific research are pervasive across disciplines and fields.

## 4.2 Understanding the temporal trends in AI impact

We seek to understand the extent to which the evolving trends in AI impact are associated with changes in AI capabilities or fields. Here, we calculate a new AI impact score at $t_2$ by holding AI n-grams and AI capabilities constant at $t_1$. Specifically, we extract AI n-grams from AI-related papers published before $t_1$ and normalize them to construct the AI n-gram frequency vector as a measure of cumulative AI capabilities. We use this vector without changing any of its AI n-grams or their frequency to calculate direct AI impact scores for the subsequent years between $t_1$ and $t_2$. In other words, only AI terms that existed before $t_1$ are considered; no new terms are added to the AI capability vector after $t_1$, and each AI term's frequency remains unchanged after $t_1$. In this way, we can separate the increase in the AI impact score into two parts: the increase due to new AI capabilities and the increase due to old AI capabilities (Supplementary Figure 9a). Specifically, the new AI part (X) is the increase in the score compared with the new score at $t_2$, while the old AI part (Y) is the increase in the new score at $t_2$ compared with the score at $t_1$ (Z).

We start by considering the period from 2015 ($t_1$) to 2019 ($t_2$). For the direct AI impact, we find that while the increase in 2019 due to old AI is not very significantly correlated with the score in 2015 (Supplementary Figure 9b), the increase in 2019 due to new AI can be largely predicted by the score in 2015 (Supplementary Figure 9c). This finding suggests that disciplines with a larger direct AI impact tend to be affected more by new AI capabilities than by old AI capabilities. In relative terms, the direct and potential AI impact scores are about 20% and 5%, respectively, above the corresponding new scores in 2019 (Supplementary Figure 7, dashed lines). The sharp increases in the direct AI impact score are remarkable (over 20%) not only in computer science, but also in many other disciplines, including physics, medicine, biology, and sociology. Together, the results suggest that disciplines are disproportionately impacted by new AI advances.

### 4.3 The relationship between potential and direct AI impact

We study the relationship between potential and direct AI impact. Specifically, we focus on one snapshot in time and explore the discrepancy between the potential and the direct AI impact of each field in 2019. To do this, we first fit a regression line and calculate the distance from each field to the average (i.e., the predicted direct AI impact by its potential). We then rank all fields according to this distance and construct three field groups (Supplementary Figure 10a): (a) "higher-direct-impact" fields are those that have higher direct AI impact than potential AI impact; (b) "higher-potential-impact" fields are those that see higher potential AI impact than direct AI impact; and (c) "others" fields are those that are not included in the first two groups.

We perform analyses to understand the features associated with the "higher-direct-impact" and "higher-potential-impact" fields. First, we study research collaboration on AI (see Supplementary Note 6 for details). Specifically, we calculate the share of each field's AI papers that are published collaboratively by domain experts and primary AI authors (i.e., collaborative AI papers). We find that the share of collaborative AI papers is higher in the "higher-direct-impact" fields than in the "higher-potential-impact" fields (two-sided Student's *t*-test; *P*-value < 0.001; Supplementary Figure 10b). Second, we consider university education in AI (see Supplementary Note 5 for details). Specifically, we estimate each field's AI education level based on its university course syllabus data. We find that the "higher-direct-impact" fields tend to be associated with a higher AI education level on average than the "higher-potential-impact" fields (two-sided Student's *t*-test; *P*-value < 0.001; Supplementary Figure 10c). Note that these analyses are correlational by nature and hence don't imply causal relationships. These results nevertheless offer support for the hypothesis that fewer collaborations with AI researchers and less education in AI may be barriers to the use of AI in unrealized fields that could have seen larger direct AI benefits.

Next, instead of the snapshot analysis, we examine the "momentum" of fields by studying the research fields that shifted from being in the "higher-potential-impact" group in 2000 to being in the "higher-direct-impact" group in 2019 (i.e., "upshift" group), and vice versa (i.e., "downshift" group). Here again, we find that fields that experienced upward shifts tended to feature more AI collaborations and higher education levels in AI (two-sided Student's *t*-test; *P*-value < 0.001 for both AI collaboration and education; Supplementary Figure 11), again consistent with the findings above. While far from being conclusive due to data limitations, these initial analyses are consistent with the hypotheses that the barriers to AI adoption may revolve around collaborations with AI experts and university education in AI. The relationship between the potential and the direct AI impact of a field is an intriguing research direction that merits substantial further investigation.

## Supplementary Note 5. Measuring AI education levels

### 5.1 Crosswalks from OSP data to MAG data

To estimate the levels of AI education in disciplines and correlate them with the AI impact, we must first align educational disciplines in course syllabi with scientific disciplines in publications and then identify AI publications referenced by course syllabus documents. Therefore, we begin by manually cross-walking the Open Syllabus Project (OSP) data to the Microsoft Academic Graph (MAG) data at the academic discipline level (from education disciplines in OSP to research disciplines in MAG) and at the scientific publication level (from publications referenced by OSP syllabi to publications in the MAG dataset), respectively.

First, we crosswalk from OSP educational disciplines to MAG research disciplines. Specifically, OSP uses two classification systems to categorize syllabus documents [59]. The first system is the Classification of Instructional Programs (CIP) system, which is a taxonomy of academic disciplines at institutions of higher education in the United States and Canada [60]. OSP uses the 2010 CIP taxonomy to determine the field that is best associated with each syllabus document. The CIP codes come in lengths of two-, four- and six-digits, where two-digit codes represent a discipline, and six-digit codes represent a more granular field of that discipline. For example, the two-digit CIP code "11" is the discipline for all "Computer and information sciences and support services" courses, and the six-digit CIP code "11.0104" is the field for all "Informatics" courses. Each syllabus document is associated with one or more CIP codes at different levels that often belong to the same two-digit CIP code category. For those with multiple two-digit codes, we only keep the first for simplicity. We manually crosswalk the two-digit CIP codes covered in the OSP data (31 codes) to the L0-level disciplines covered in the MAG data (14 disciplines) by matching the names of these disciplines and selecting the closest match for each CIP code.

Another classification system is the OSP field name classifier, which is trained on a subset of the CIP taxonomy for describing syllabi [59]. The OSP classification system generally combines CIP codes within the same two-digit branch of the taxonomy into a field name. For example, CIP code "45.09" for "International Relations and National Security Studies" and CIP code "45.10" for "Political Science and Government" are both under CIP code "45" for "Social Sciences" and are combined into the OSP field named "Political Science." We manually crosswalk OSP field names (62 names) to MAG disciplines/fields (17 L0-level disciplines and 45 L0/L1-level fields) by matching field names and selecting the closest match for each OSP field name. Further, we combine these two classification systems to assign only one MAG L0-level discipline (or field at the L1 level) to each syllabus document that best describes its academic discipline. We rely on the OSP field name classification system and supplement it with the CIP classification system when there are missing values in the CIP codes or inconsistencies in the matched L0-level disciplines.

Second, we match publications cited by OSP syllabus documents to scientific publications in MAG. Specifically, for publications with DOI information, we use DOI to match them to MAG publications directly. To reduce false positives, we confirm that the last names of the first author for each matched publication record are the same. For publications without DOI information, which are typically books and conference papers, we match them to MAG publications based on the publication's title, the last name of the first author, and the publication year. We exactly match the title and last name but allow a year difference in the publication year. In total, we match about 924K publications cited by syllabus documents published between 2000 and 2018.

## 5.2 Estimating the levels of AI education

Based on the linkages from OSP to MAG data at the discipline level and the publication level, we estimate the levels of AI education in each educational discipline. We assume that an educational discipline's AI education level is higher if syllabus documents in the discipline cite more publications related to AI because these citations are a signal that the course teaches more AI knowledge. For each educational discipline, we calculate the share of citations from syllabus documents in the discipline to AI publications identified from the MAG data using the five AI field categories, and we use this measure as the level of AI education for the discipline.

Before we calculate the levels of AI education based on the OSP data, we filter the syllabus documents to ensure that we are capturing meaningful measures of AI education. First, we use syllabus documents that cite at least five scientific references to proxy courses taught at the graduate level and above. The reasoning here is that syllabi with fewer references may more likely correspond to undergraduate-level courses, and those with more references may correspond to graduate-level or more research-oriented courses. Students who take these higher-level courses should have a higher propensity to perform research and thus better represent the student population that benefits from AI education. Second, we only use syllabus documents that were published in the most recent five-year period, 2014-2018, according to our OSP dataset. As we find that the impact of AI is most widespread since 2015, measuring AI education levels based on recent syllabi is more relevant. Alternatively, we could use other criteria to filter the syllabus documents or estimate AI education levels using other measures, including the share of syllabus documents that cite at least one AI publication. These results are robust across several analytical choices (see Supplementary Note 5.3 for details).

**5.3 Relationships between AI education and AI impact**

Analyzing the relationship between the level of AI education and the impact of AI on scientific research, we find that the discipline-level correlation decreases, and so does its significance after excluding the top three computation-intensive disciplines from the analysis (i.e., computer science, mathematics, and engineering). This observation holds robustness for both the direct AI impact and the potential AI impact that we estimate based on publication and patent data. Overall, our analyses convey three sets of findings across several measures. First, we find that the discipline-level correlation between AI education levels and AI impact scores is insignificant after excluding the top three disciplines (Supplementary Figure 12a). Second, we compare three samples: syllabi with one to five references, syllabi with at least five references, and syllabi with at least ten references. After repeating our analyses for these three samples, we find that consistent with our intuition, the correlation between AI education and AI impact is weaker for the group with one to five references than the group with at least five references (Supplementary Figure 12b). We also find that our results remain robust when using data with at least ten references to identify graduate-level courses (Supplementary Figure 12c). Third, we find that the results are robust when using the share of syllabi that cite at least one AI publication to estimate AI education levels (Supplementary Figure 12d), and the correlation is weaker in the earlier periods of 2008-2013 and 2000-2007 compared to 2014-2018 (Supplementary Figures 12e,f).

## Supplementary Note 6. Measuring collaborations on AI

### 6.1 Identification of AI researchers and domain experts

Based on AI papers in disciplines other than computer science, we estimate the level of cross-discipline collaboration on AI-related research between domain experts and AI researchers. First, we assign a primary discipline to each researcher. Each paper belongs to one primary discipline in our data, so we count each author's publications in each discipline and assign the discipline in which they published most frequently in the period 1960-2019 to be the author's primary discipline. If an author has multiple most frequent disciplines, we randomly select one discipline. Second, we treat authors whose primary discipline is computer science (CS) as computer scientists and those with primary disciplines other than CS as domain experts. Although authors who frequently publish CS papers may specialize in general computer science and not necessarily in AI research, we use computer scientists as an imperfect proxy for identifying AI researchers and further check its robustness using an alternative approach (see Supplementary Note 6.3 for details). Third, based on the composition of authors' primary disciplines, we categorize each AI paper published in a discipline other than CS into one of the four co-authorship types: (1) "domain & CS," which involves both domain experts and computer scientists; (2) "domain sole," which involves only domain experts; (3) "CS sole," which involves only computer scientists; and (4) "others," which involve neither domain experts nor computer scientists (e.g., an AI paper in the geology discipline published by two physicists who are not domain experts according to our identification method). Then, we calculate the share of collaborative AI papers (i.e., the "domain & CS" co-authorship type) for each discipline and correlate it with the AI impact measures.

### 6.2 Temporal trends in cross-discipline AI collaboration

We analyze cross-discipline collaboration patterns for AI-related papers co-authored by domain experts and/or computer scientists. Here we consider AI papers published by at least two authors in the period 1980-2019 and major disciplines that have the largest AI impact scores and the highest AI education levels. We find that while both the number of collaborative AI papers (i.e., "domain & CS") and the number of domain-only AI papers (i.e., "domain sole") increase over time (Supplementary Figures 13a,b), the growth rate of collaborative AI papers is larger than that of domain-only AI papers (Supplementary Figure 13c). Hence, in terms of the relative share, we see that the share of domain-only AI papers decreases over time across various disciplines (Supplementary Figure 13d). Meanwhile, there is an increasing share of collaborative AI papers in major computation-intensive disciplines, especially in recent years (Supplementary Figure 14). In particular, the share of collaborative AI papers increases rapidly in engineering and mathematics, suggesting a growing reliance on AI expertise by domain experts or their increasing contribution to AI research. Together, these results indicate that teamwork and cross-domain collaboration may be particularly important as AI's impact on research becomes more pervasive in the future.

**6.3 Robustness checks on the identification of AI researchers**

To test the robustness of the AI collaboration results, we considered alternative approaches to identifying AI researchers, as well. An author's career titles, such as the discipline in which they obtained a Ph.D. or their academic department, are most informative for identifying primary disciplines. Unfortunately, the MAG dataset does not include such information for each author, and we are unable to identify AI researchers by applying AI terms to authors' titles. However, we develop an alternative approach to distinguishing primary AI authors based on AI publications in the MAG data. Specifically, we identify primary AI authors as those who published at least three AI papers according to the MAG five AI field categories and for whom more than half of their publications are AI papers—in other words, those who frequently publish AI papers [36]. To identify a primary discipline for each author, we use our previous approach, which is based on the discipline in which they publish most frequently (see Supplementary Note 6.2 for detailed explanations). This new method allows us to identify AI researchers based on their publications that are directly relevant to AI, which is relatively more precise than the proxy of computer scientists in general. We show that our main findings are robust with this new method of AI researcher identification (Supplementary Figure 15).

## Supplementary Note 7. Measuring demographic disparities

### 7.1 Crosswalks from SDR data to MAG data

We infer the gender and race/ethnicity compositions of each discipline based on the 2017 Survey of Doctorate Recipients (SDR) data, which provides summary statistics on demographic information (see Supplementary Note 1.4). The SDR field of doctorate taxonomy contains field categories at different levels, including three categories at the top level ("Science," "Engineering," and "Health"), as well as categories at a granular level (e.g., "Economics" and "Physics"). This taxonomy is aligned with the National Center for Science and Engineering Statistics (NCSES) Taxonomy of Disciplines (ToD). We manually crosswalk the 31 SDR fields to the 15 MAG disciplines by comparing the names of these fields and disciplines. We match each SDR field to the single MAG discipline that best represents its scientific discipline. The SDR data also report statistics at a more granular field level. For simplicity, we only crosswalk at the discipline level as we focus on overall trends. Further, we aggregate the SDR data to each matched MAG discipline to calculate the discipline's gender and race/ethnicity composition.

### 7.2 Gender disparity in AI impact

We quantify gender disparity using two measures. First, we calculate the share of women researchers in each discipline. We find a negative correlation between the share of women researchers and the AI impact score for both the direct and potential impact (Figs. 4a,b in the main text). Second, we aggregate AI scores for each gender group by weighting the discipline-level AI scores using the gender share of disciplines. A larger average AI score for a gender group indicates that researchers in the group have a greater AI impact and thus benefit more from AI. We find that women researchers, on average, benefit less from AI (Figs. 4c,d in the main text) than men researchers. These observations suggest a gender gap across disciplines regarding AI benefit.

### 7.3 Race disparity in AI impact

The SDR data also report the number of researchers in each race group, including "white," "Asian," and underrepresented minorities (URM). The URM group comprises "Black," "American Indian or Alaska Native," "Hispanic or Latino," and "Other race." We quantify racial disparity using two measures. First, we calculate the share of URM researchers in each discipline. We find a negative correlation between the share of URM researchers and the AI impact score (Figs. 4e,f in the main text). Second, we aggregate AI scores for each racial group by weighting the discipline-level AI scores using the race share of disciplines. We find that URM researchers receive the least AI benefits, especially Black researchers (Figs. 4g,h in the main text). The literature suggests that workers of color are particularly vulnerable to job loss due to automation and AI [61, 62], and our results suggest they are also less likely to benefit from AI in scientific research.

# Supplementary Note 8. Robustness check of main results

## 8.1 Precision-recall analysis of identifying AI research

We analyzed the precision and recall of our approach in identifying AI research (see Supplementary Note 2.1 for details). First, we randomly sampled 100 AI papers, defined as those that mention at least one AI-related n-gram, regardless of the weight or importance of the particular n-gram(s) in representing AI research. We then manually determine if these papers actually use AI methods. We find that 14 of the papers fall into one of the MAG five AI fields ("machine learning," "artificial intelligence," "computer vision," "natural language processing," and "pattern recognition"), 69 papers are AI application papers, and 17 papers appear to be false positives that mention less representative and less common AI n-grams. Therefore, we estimate that the approach has a precision of around 0.83. Second, we randomly sample 100 AI papers, defined as those that are identified by one of the four arXiv "Subject Categories": cs.AI (Computer science: Artificial Intelligence), cs.LG (Computer science: Machine Learning), cs.NE (Computer science: Neural and Evolutionary Computing), and stat.ML (Statistics: Machine Learning) [30-33], and we determine if they are covered by our n-gram approach. We find that 55 papers fall into one of the five MAG AI fields, 33 papers mention at least one AI n-gram, and 12 papers appear to be false negatives. We estimate, therefore, that the approach has a recall of around 0.87. Thus, the method has a high level of precision and recall, and it balances these two metrics well.

We test the robustness of our findings using two alternative approaches to defining AI publications. In the first approach, we identify the top 20 most frequent AI n-grams (Supplementary Table 2) and extend the five AI fields to these 20 n-grams across all "field of study" categories, from the L1 level to the L5 level in MAG (see Supplementary Note 1.1 for details). We treat publications in any of these 20 fields as AI publications. In the second approach, we identify AI publications using "Subject Categories" in the arXiv dataset (2023-11 version) [63] which provides the metadata of each preprint. There, we matched arXiv preprints to MAG publications using the arXiv Identifier. Following previous works [30-33], we treat preprints in four arXiv "Subject Categories" as AI publications: cs.AI, cs.LG, cs.NE, and stat.ML. We apply our frameworks to AI publications defined by these two alternative approaches and find that the main findings are largely robust (Supplementary Figure 16). The literature also suggests other ways to identify AI publications, including, for example, topic modeling and embedding-based methods [64, 65].

## 8.2 Robustness test on the direct AI impact calculation

The calculation of the direct AI impact score assumes that AI has a large impact on field X if important AI techniques are frequently mentioned by papers in the field. It is essential to recognize that the landscape of AI research evolves rapidly, with some AI methods becoming more popular and some less popular over time, and this is why we distinguish AI n-grams by their level of popularity. We use the dot product because it is a simple measure that balances the importance of AI n-grams and their frequency of use in a field. In the AI vector, the value of each n-gram coordinate signals its importance in representing AI research. In the field vector, the value of each n-gram coordinate indicates its frequency of use in the field. The dot product helps to capture both the importance of AI n-grams and their frequency of use in each field.

We check the robustness of our results using two alternative measures. In the first measure, we treat each n-gram with the same importance in the AI vector and calculate the sum of the field vector for its AI n-gram coordinates. Formally, $S_D^t = \sum \hat{G}_B^t \cdot I_{AI}$, where $I_{AI}$ represents an all-one vector for AI n-grams. This sum measure captures the total frequency of AI n-grams in a field. In the second measure, we calculate the weighted average for the field vector using the AI vector as the weight for AI n-grams. Formally, $S_D^t = \sum \hat{G}_B^t \cdot \hat{G}_{AI}^t / \sum \hat{G}_{AI}^t$. This weighted average measure captures the average frequency of AI n-grams in a field. We find that the results of these two alternative measures are highly correlated with the results of the dot product measure (Supplementary Figure 17), suggesting the robustness of the main findings.

## 8.3 Robustness test on the potential AI impact calculation

We validated the "AI capability–field task framework" using several approaches. First, we use the future of work literature, which has estimated the impact of AI on labor [26, 47-53], as the foundation for our measurements. Specifically, this literature has used the texts of patents to infer AI capabilities, used the descriptions of occupations to infer job tasks, and measured the exposure of jobs to AI using the alignment between AI capabilities and job tasks [47]. We build on this literature to estimate the potential AI impact on science. Second, we devised two alternative methods to calculate the potential AI impact score. In the first method, we calculate the fraction of AI verb-noun pairs in each research field, treating each AI verb-noun pair with the same weight. Formally, the fraction measure is defined as $S_P^t = \sum \hat{C}_B^t \cdot I_{AI}$, where $\hat{C}_B^t$ represents the field verb-noun vector, and $I_{AI}$ represents an all-one vector for the AI verb-noun vector. In the second method, we calculate the potential AI impact score using each field as the baseline, assuming that field tasks are well aligned with field capabilities. Formally, the field baseline measure is defined as $S_P^t = \sum \hat{C}_B^t \cdot \hat{C}_{AI}^t / \sum \hat{C}_B^t \cdot \hat{C}_B^t$. We find that the potential AI scores are highly correlated (Pearson's correlation 0.95 and 0.90, respectively), suggesting the robustness of these measurements.

## Supplementary Note 9. Career effects of engaging in AI research

### 9.1 Name-based demographic data inference

We study the possibility that engaging in AI research could lead to career-changing experiences for women and URM researchers. This requires us to perform career-level analyses. To do this, we employ naming-based inference algorithms (nomquamgender [66, 67] and demographicx [68, 69]) to infer the gender and race of each author in MAG, although the limitations of these algorithms are noted (see Supplementary Note 1.4 for details). We only consider authors whose gender can be confidently inferred to be "woman" or "man" for the gender analysis and whose race can be confidently inferred to be "white," "Asian," "Black," or "Hispanic" for the race analysis. Based on this group of authors, we then construct a sample of authors who published at least ten papers from 2000 to 2019 and have at least five years of career history (i.e., the time difference between their first and last papers). We identify AI researchers as those who published at least three AI papers as AI researchers and pinpoint the start of their AI engagement by tracing their first AI paper.

### 9.2 Effects on citation premium and leading the field

We calculate the average hit rate of papers by an author before and after engaging in AI research, measuring the fraction of papers that land within the top 5% of papers ranked by total citations in the same field and year. We find that, on aggregate, the average hit rate of papers by a researcher exhibits an increase right after engaging in AI research (Supplementary Figure 18a), highlighting the citation premium of AI engagement. However, when we further unpack these results by gender and race, we find that the citation premium is more concentrated in the overrepresented groups, and women and URM researchers appear to benefit less from AI engagement than their counterparts (Supplementary Figures 18b,c). This opens up a new question for further research.

We also examined whether women or URM authors lead the way in their fields in collaborating with AI researchers. Here we assign a primary discipline to each author based on the discipline in which the individual publishes most frequently and treat those with a primary discipline of computer science as AI researchers (see Supplementary Note 6.1 for detailed explanations). We identify the top 5% of researchers in each field in terms of 1) the number of papers and 2) their number of hit papers as those who lead the way in their respective fields. We find that while researchers who collaborate with AI researchers indeed tend to lead the way in their respective fields (Supplementary Figures 19a,d), the effect is more salient for men and white researchers (see Supplementary Figures 19b,e for gender and Supplementary Figures 19c,f for race). The results are robust when we use alternative methods to identify AI researchers, as mentioned above (see Supplementary Note 6.3 for details). Together, these results provide some preliminary evidence that women and URM researchers may see fewer benefits from AI advances.

## Supplementary Figures

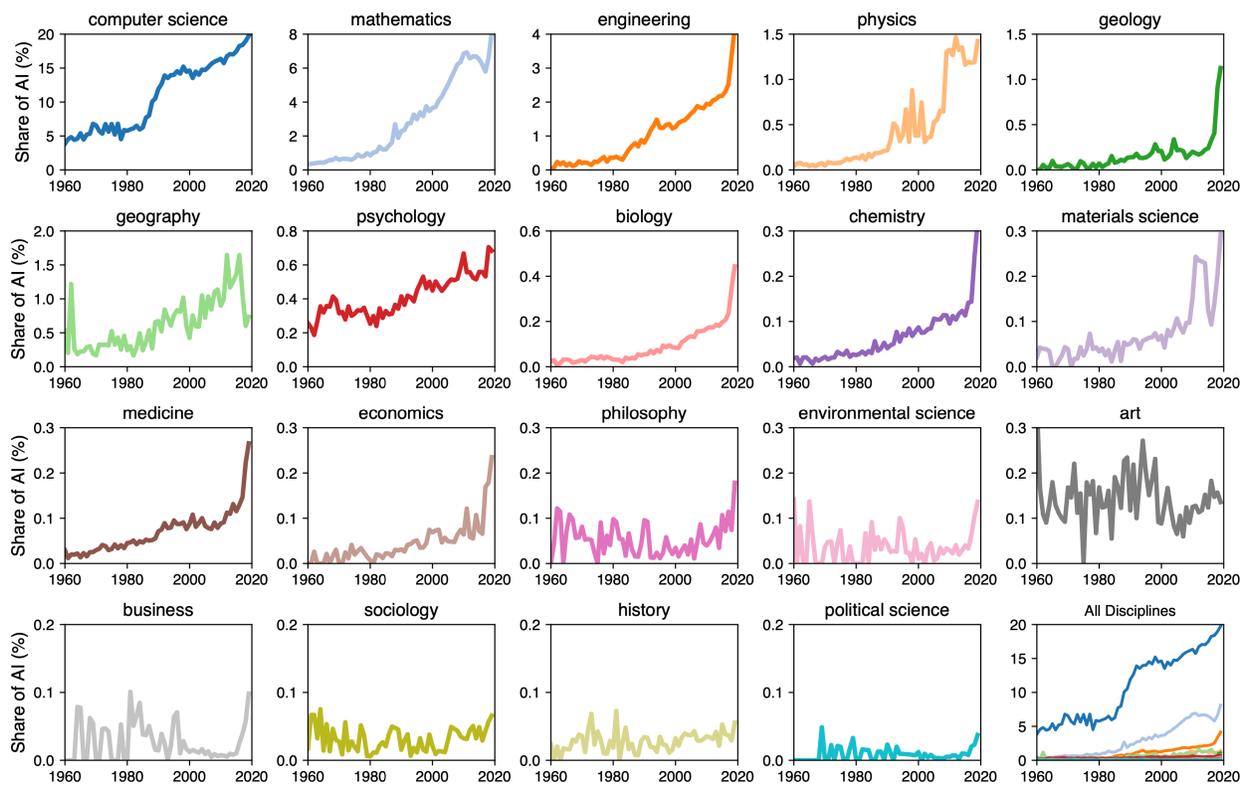

**Supplementary Figure 1. The share of AI papers in each discipline.** Each panel corresponds to a discipline, and disciplines are ordered by their share of AI papers in 2019, from the largest share (upper left) to the smallest share (lower right). The lower-right panel compares different disciplines in the same plot with the same y-axis.

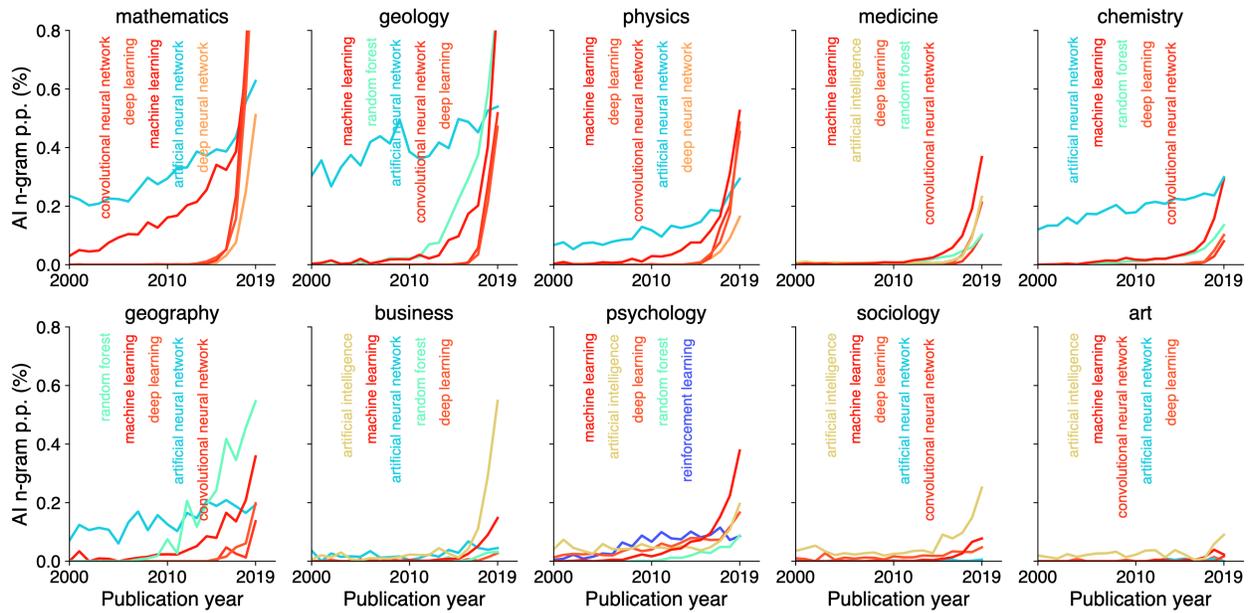

**Supplementary Figure 2. AI n-grams and temporal trends in each discipline.** Top 5 AI n-grams per paper for a few major disciplines. These AI n-grams are extracted from papers published in each discipline in the period 2000-2019, and they are colored according to their relative rankings in AI papers. The color code is the same as the one in Fig. 1c of the main text.

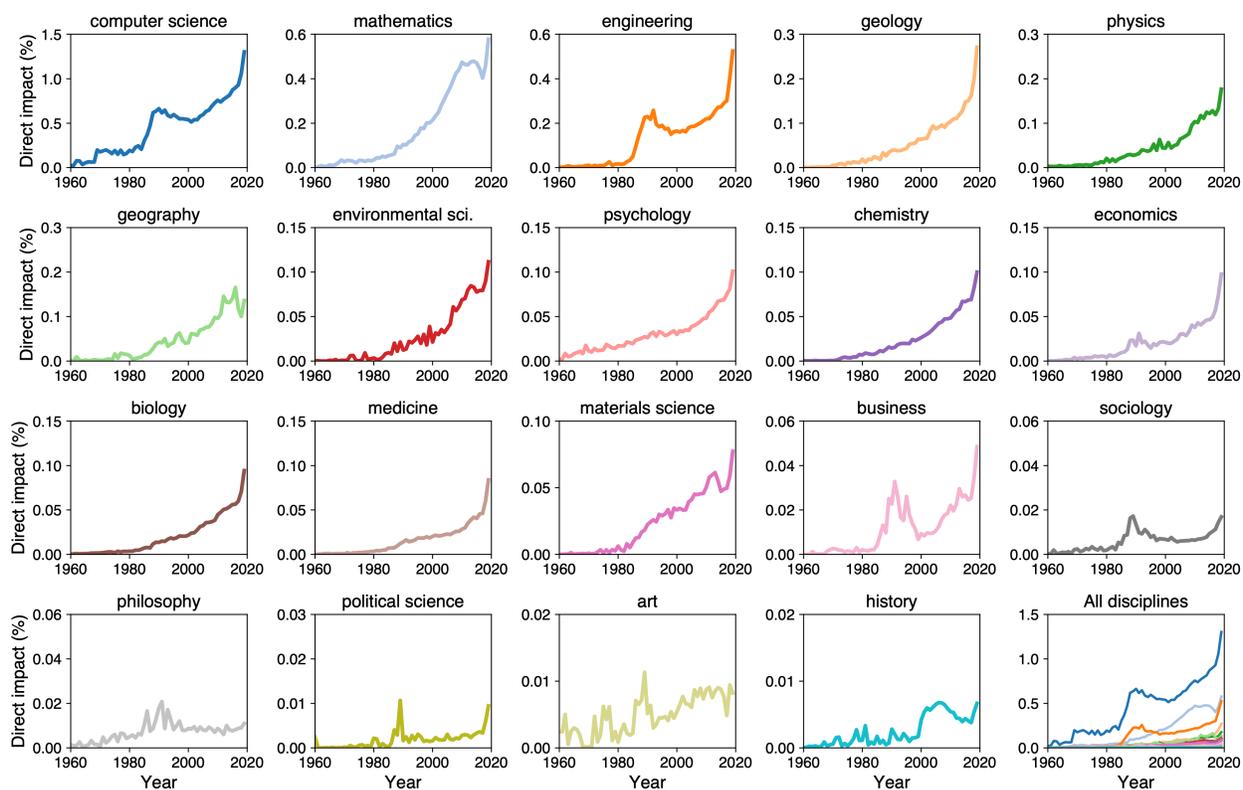

**Supplementary Figure 3. Temporal trends in the direct AI impact score for each scientific discipline.** The results cover the period between 1960 and 2019. Each panel corresponds to a discipline, and disciplines are ordered by their direct AI impact scores in 2019, from largest impact (upper left) to smallest impact (lower right). The panel in the lower-right corner compares the direct AI impact score across different disciplines. All curves are under the same y-axis, and the color of each curve corresponds to the one in each panel.

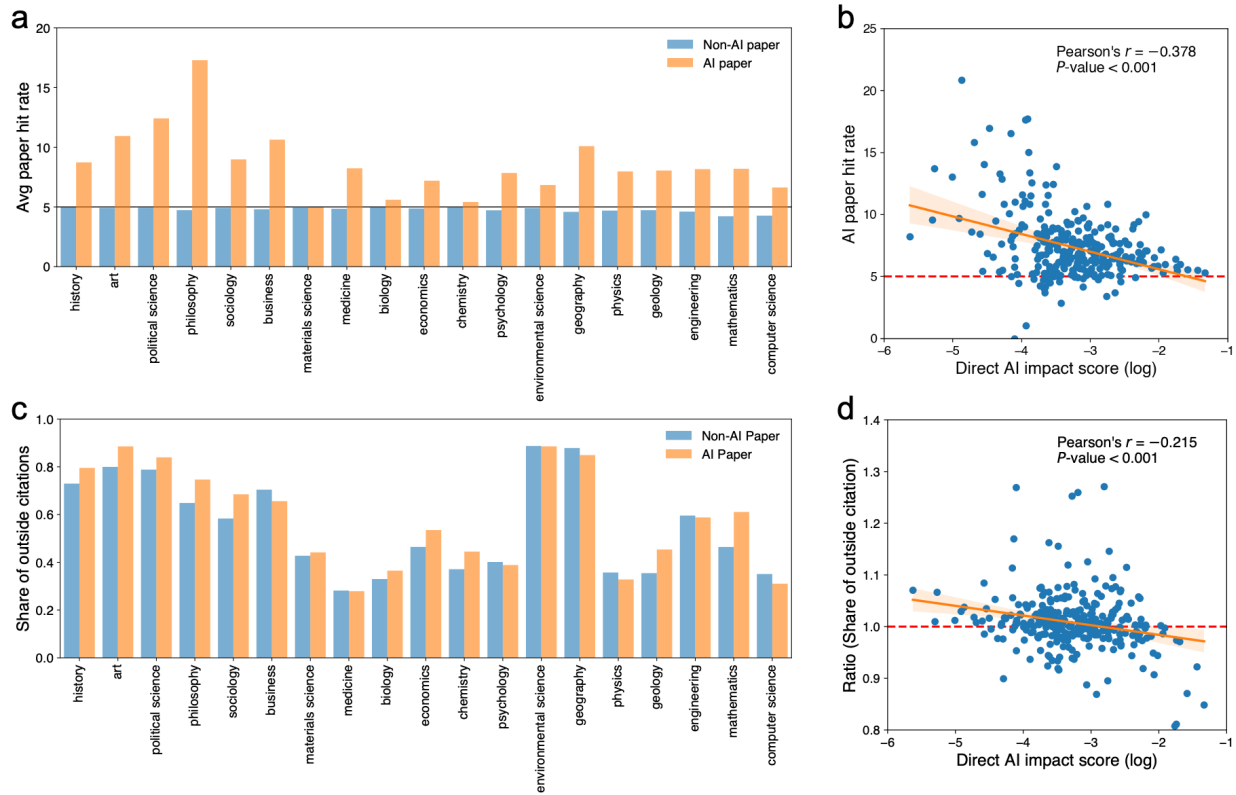

**Supplementary Figure 4. Citation impact of papers that mention AI-related terms. (a)** The rate of AI papers (i.e., those that mention at least one AI-related term) being hit papers and the same rate for non-AI papers (i.e., those that do not mention any AI-related terms) in each research discipline. Hit papers are defined as the top 5% of papers ranked by total citations in the same field and same year. Disciplines are presented in ascending order based on their direct AI impact scores. The horizontal line marks the 5% hit rate. **(b)** The negative correlation between the direct AI impact score and the hit paper rate for AI papers in each field. The correlation was determined using a two-sided Pearson's correlation test. Linear fit (centre line) of the data with 95% confidence intervals (error bands) is shown. The horizontal dashed line marks the 5% hit rate. **(c)** The share of all citations that both AI and non-AI papers receive in each discipline that are outside-discipline citations. Disciplines are presented in descending order of the share for non-AI papers. **(d)** The negative correlation between the direct AI impact score and the ratio of the share of outside-field citations in AI papers to the share of outside-field citations in non-AI papers. The correlation was determined using a two-sided Pearson's correlation test. Linear fit (centre line) of the data with 95% confidence intervals (error bands) is shown. The horizontal dashed line marks 1, the value at which AI papers receive the same share of outside-field citations as non-AI papers.

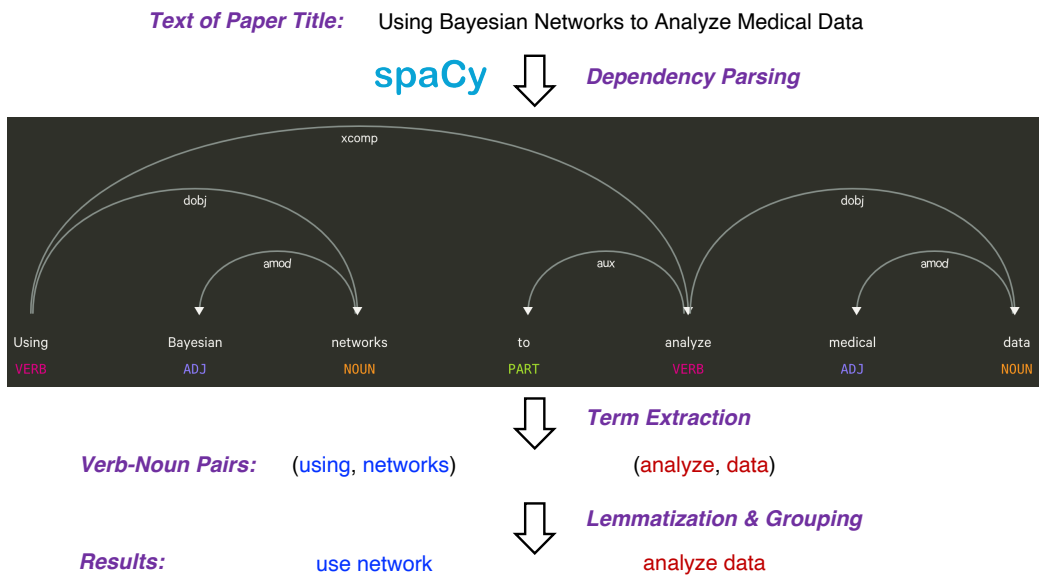

**Supplementary Figure 5. Illustration of the extraction verb-noun pairs from publications.** The analysis uses the dependency parsing algorithm embedded in spaCy, an open-source Python library for advanced NLP applications. The spaCy dependency parser is used to identify the "dobj" (direct object) relationship among words in the title of a publication, where verb-noun pairs are extracted by pairing up verb and noun words under the "dobj" relationship. Then, lemmatization and standardization are used to clean verbs and nouns, and words with similar meanings are grouped into one word. This process produces standardized verb-noun pairs.

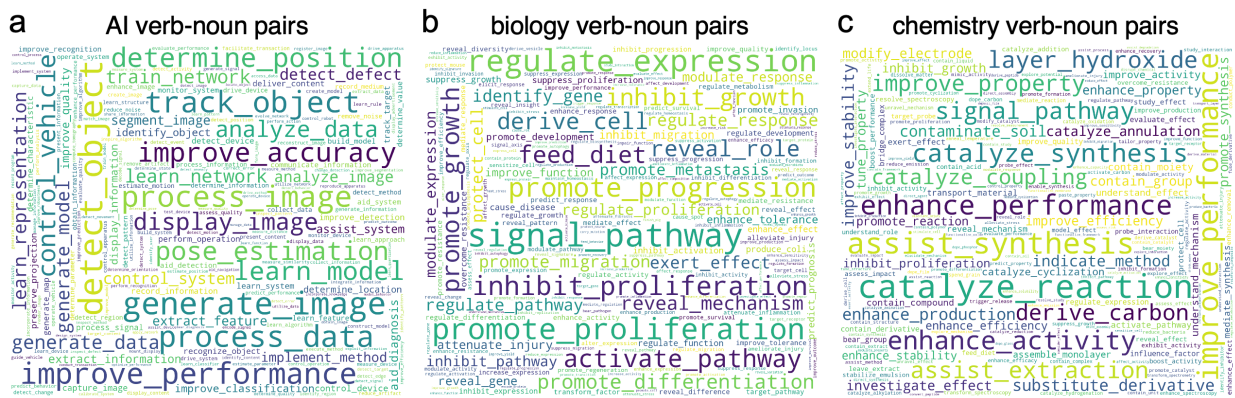

**Supplementary Figure 6. Word cloud of verb-noun pairs extracted from publications and patents. (a)** Verb-noun pairs extracted from the titles of AI-related papers and patents in the period of 1960-2019. **(b)** Verb-noun pairs extracted from the titles of biology publications in 2019. **(c)** Verb-noun pairs extracted from the titles of chemistry papers in 2019.

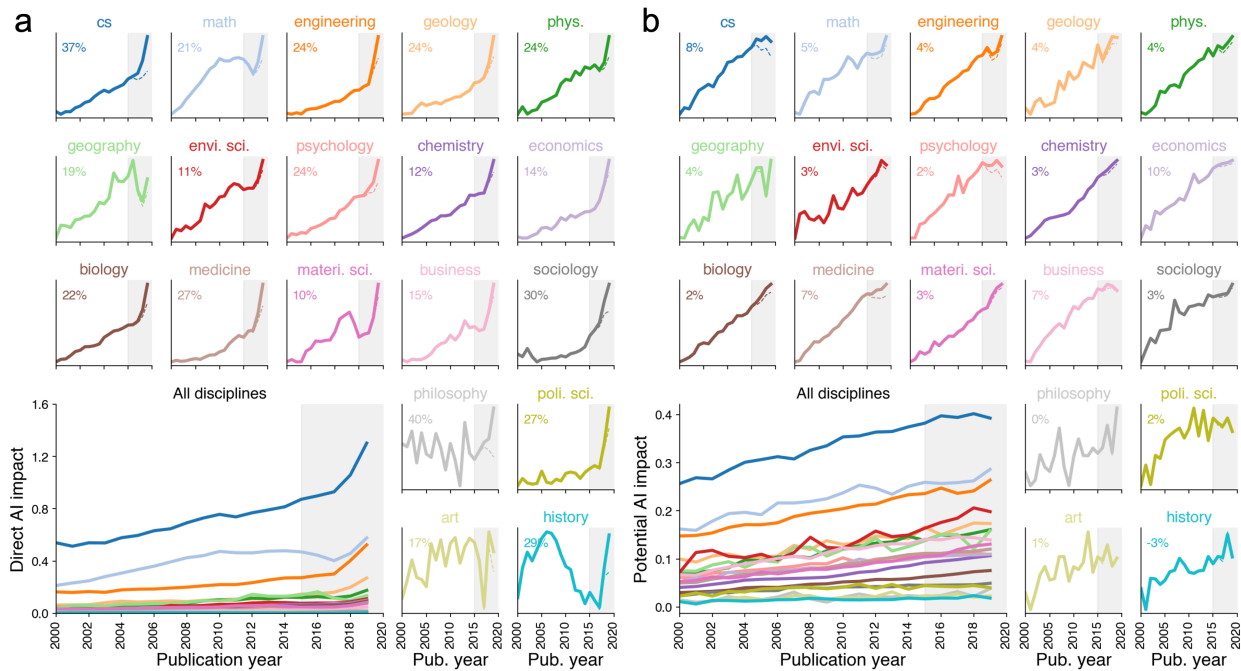

**Supplementary Figure 7. Trends in the direct and potential AI impact. (a)** Temporal trends in the direct AI impact score. The dashed line shows the new score calculated using cumulative AI n-grams fixed in 2015 and the current AI n-grams. Percentage changes in the scores are shown. Disciplines are sorted by their direct AI impact score in 2019. The lower-left panel shows all results colored by disciplines according to each panel. **(b)** Temporal trends in the potential AI impact score. The dashed line shows the new score calculated using cumulative AI capabilities fixed in 2015 and the current field tasks. The order and color of each discipline are the same as panel (a).

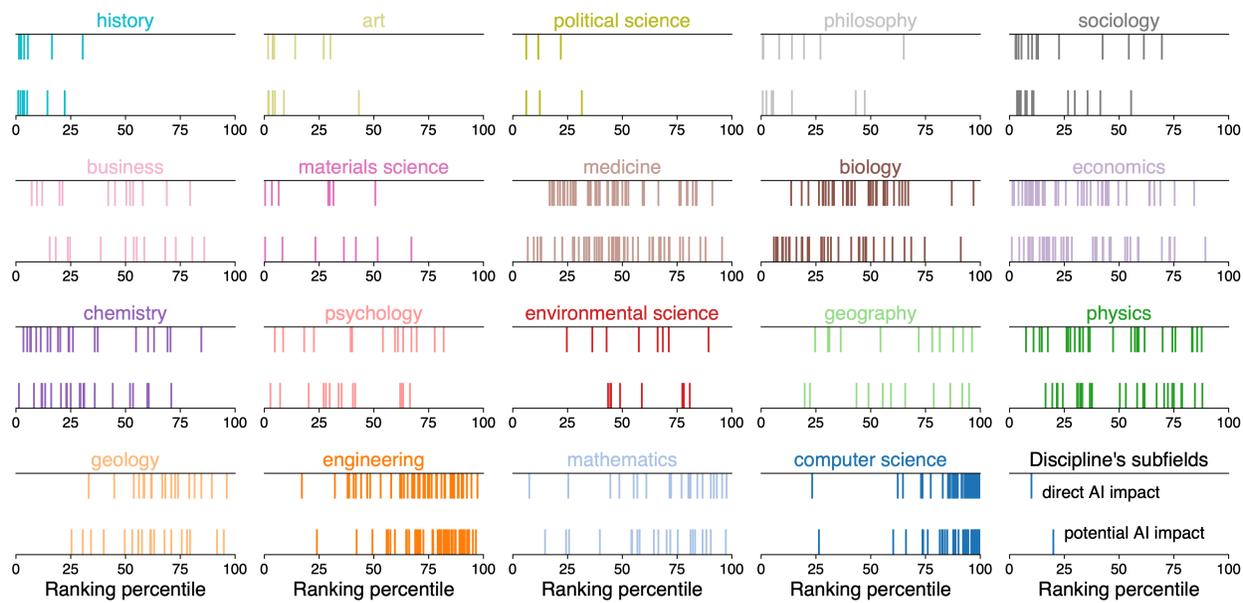

**Supplementary Figure 8. Alignment between direct and potential AI impact despite the within-discipline heterogeneity.** The plot shows the percentile of each discipline's subfields among all 292 MAG fields. The legend is presented in the bottom-right panel. Fields are ordered by the direct AI scores in the upper row and the potential AI scores in the lower row.

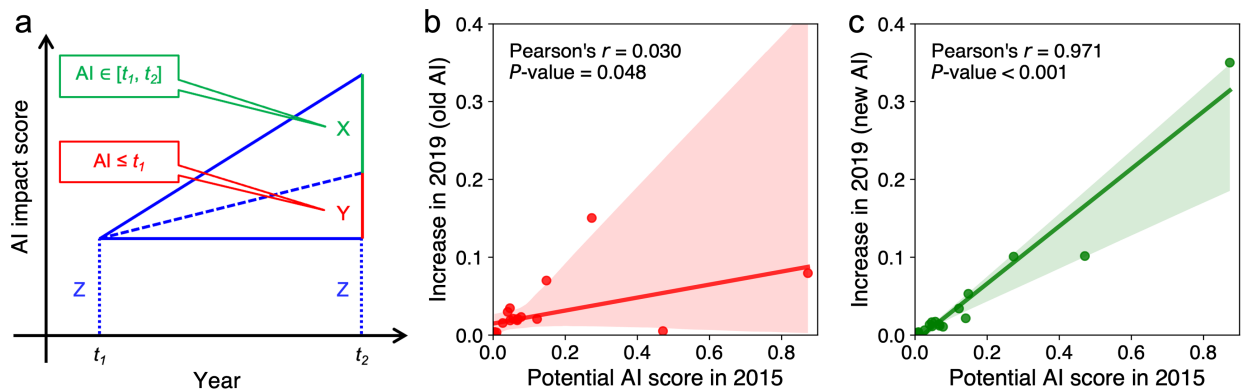

**Supplementary Figure 9. Understanding the increase in AI impact.** (a) Illustration of the decomposition of the increase in AI impact scores. Z is the score at $t_1$, Y+Z is the new score in $t_2$, and X+Y+Z is the score at $t_2$. The increase in AI impact from $t_1$ to $t_2$ consists of X (the component that results from new AI capabilities; in green) and Y (the component that results from old AI capabilities; in red). (b) The correlation between the direct AI impact score in 2015 (Z) and the increase in 2019 due to old AI capabilities (Y). The correlation was determined using a two-sided Pearson's correlation test. Linear fit (centre line) of the data with 95% confidence intervals (error bands) is shown. (c) The correlation between the direct AI impact score in 2015 (Z) and the increase in 2019 due to new AI capabilities (X). The correlation was determined using a two-sided Pearson's correlation test. Linear fit (centre line) of the data with 95% confidence intervals (error bands) is shown. All correlations were determined using a two-sided Pearson's correlation test.

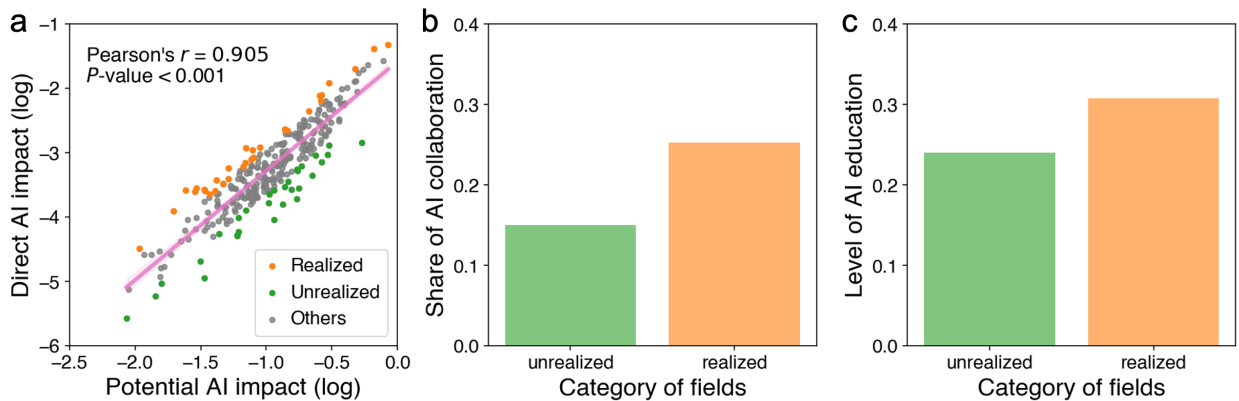

**Supplementary Figure 10. Understanding the potential vs. direct impact of AI in research fields.** **(a)** Categorizing fields in the "potential-direct" AI impact plane in 2019. The correlation was determined using a two-sided Pearson's correlation test. Linear fit (centre line) of the data with 95% confidence intervals (error bands) is shown. **(b)** The differences in the collaboration on AI between the two groups. **(c)** The differences in the education in AI between the two groups. The differences were tested using a two-sided Student's *t*-test.

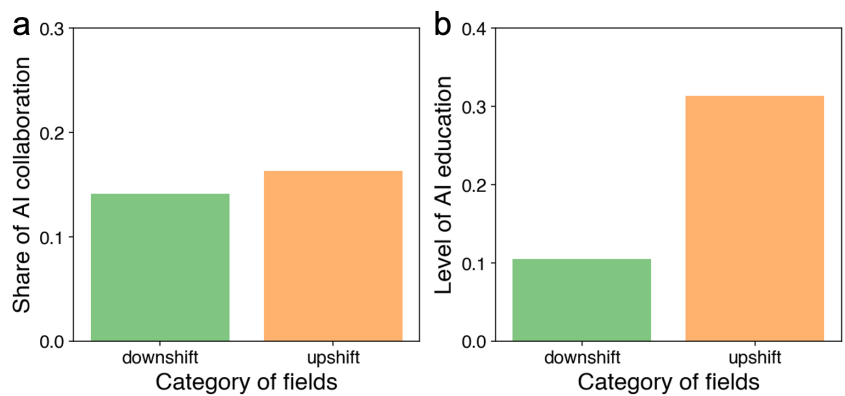

**Supplementary Figure 11. Categorizing fields using their "momentum" from 2000 to 2019.** Research fields that shifted from "higher-potential" in 2000 to "higher-direct" in 2019 are in the "upshift" group; research fields that shifted from "higher-direct" in 2000 to "higher-potential" in 2019 are in the "downshift" group. **(a)** The differences in the collaboration on AI between the two groups. **(b)** The differences in the education in AI between the two groups. The differences were tested using a two-sided Student's *t*-test.

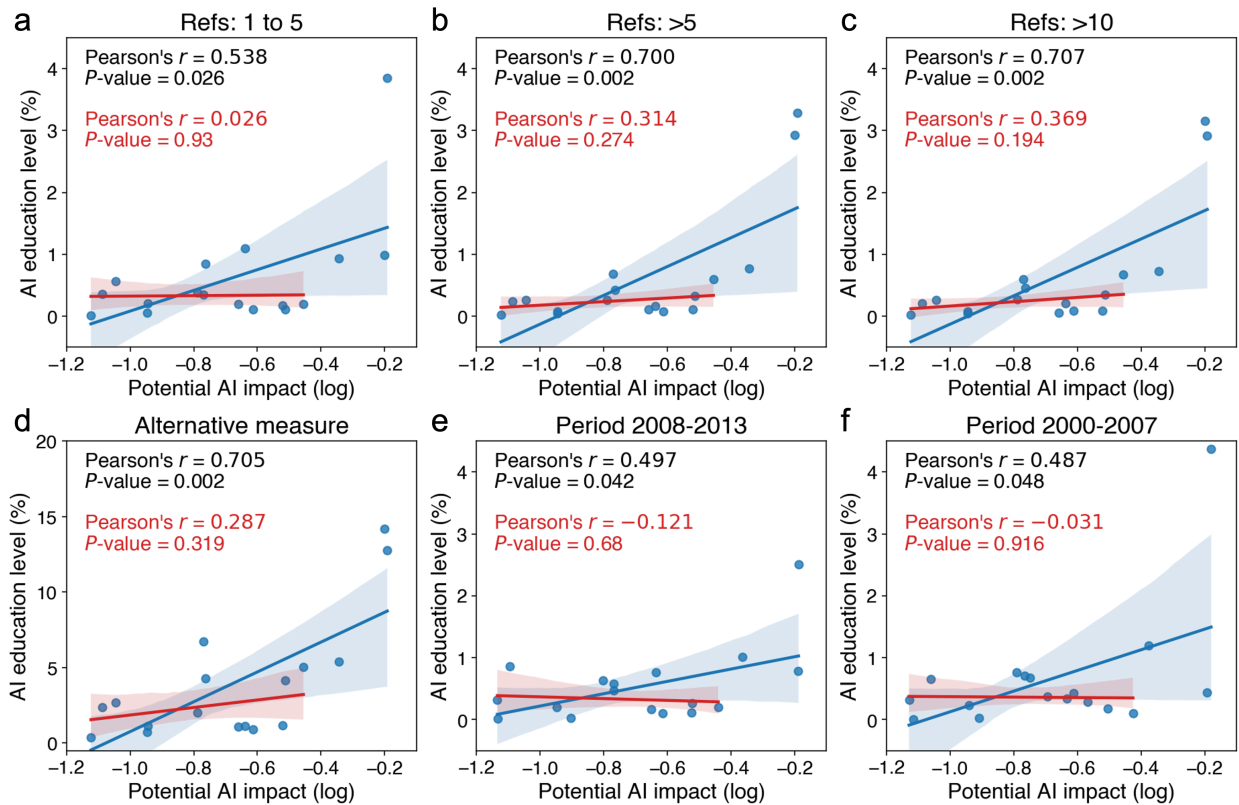

**Supplementary Figure 12. Correlations between AI education and AI impact.** The level of AI education is calculated using syllabus documents that cite **(a)** one to five scientific references, which is a proxy for undergraduate-level courses; **(b)** at least five scientific references, which is a proxy for graduate-level courses; **(c)** at least ten references, which is a proxy for more research-oriented courses. **(d)** The level of AI education is calculated by the share of syllabus documents that cited at least one AI publication. Only syllabus documents from 2014 to 2018 are used for the calculations in the panels (a-d). **(e)** The level of AI education is calculated using syllabus documents with at least five references from 2008 to 2013. **(f)** The level of AI education is calculated using syllabus documents with at least five references from 2000 to 2007. Throughout all panels, the correlation was determined using a two-sided Pearson's correlation test. Throughout all panels, linear fits (centre lines) of the data with 95% confidence intervals (error bands) are shown, and all correlations were determined using a two-sided Pearson's correlation test.

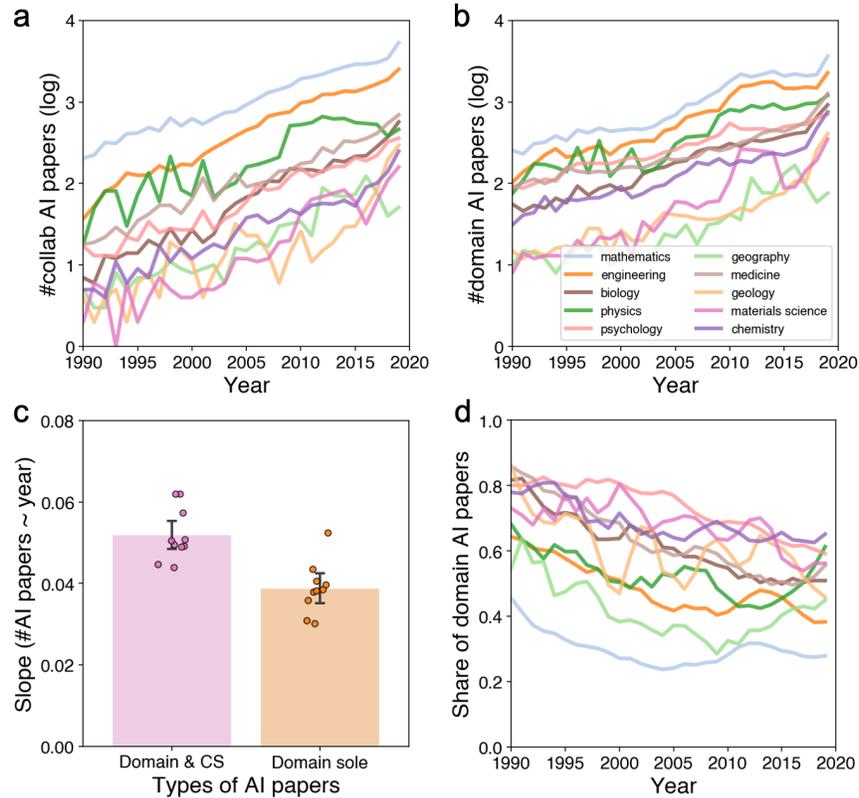

**Supplementary Figure 13. The growth in AI publications by different co-authorship types from 1990 to 2019 in each discipline.** (a) Collaborative AI papers, i.e., "domain & CS." (b) Domain AI papers, i.e., "domain sole." (c) The slope of a linear fit for AI papers. The number of "domain & CS" papers has, on average, a larger slope than "domain-only" papers. Scatters represent disciplines, bar plots show mean values, and error bars represent standard errors for the disciplines (n=10). (d) Temporal changes in the share of domain-only AI papers in disciplines.

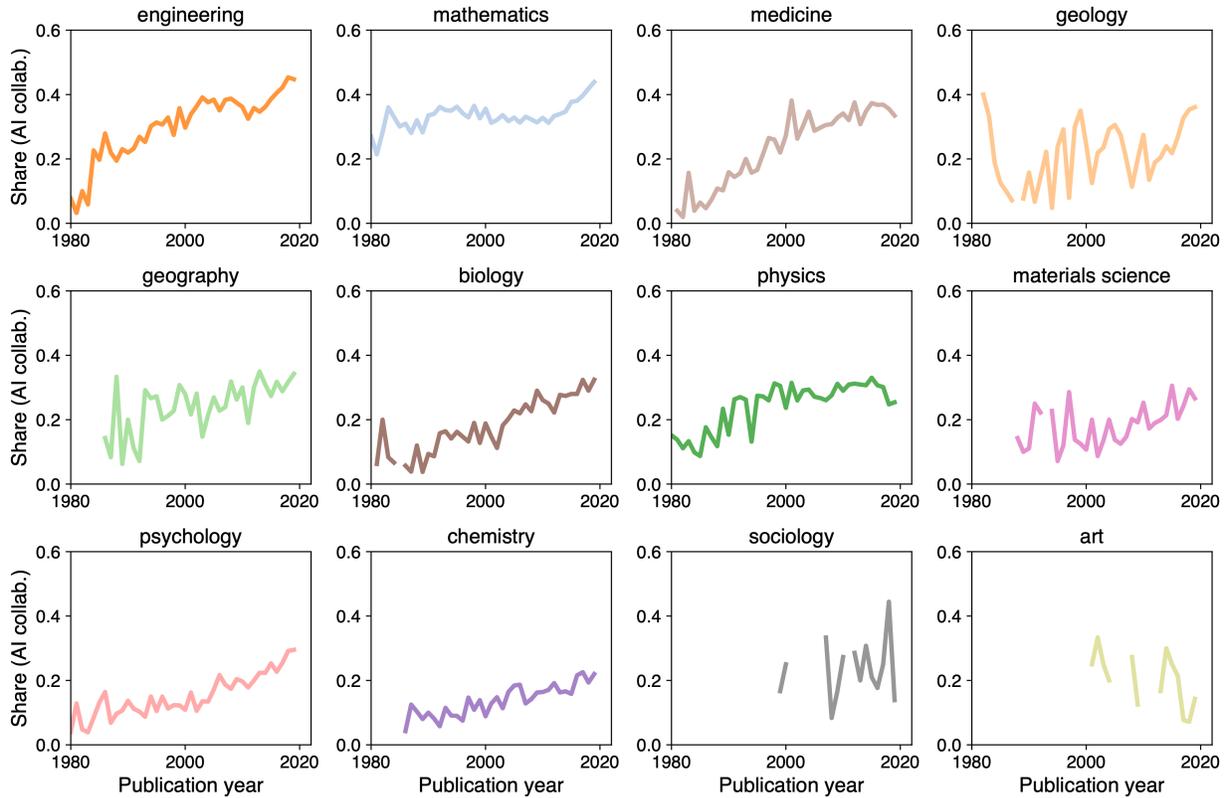

**Supplementary Figure 14. Temporal trends in cross-discipline AI collaboration.** The share of collaborative AI papers co-authored by domain experts and computer scientists (i.e., "domain & CS") in each discipline. Only AI-related papers published outside of the computer science discipline in the period 1980-2019 are involved in the calculation. Only top disciplines are shown.

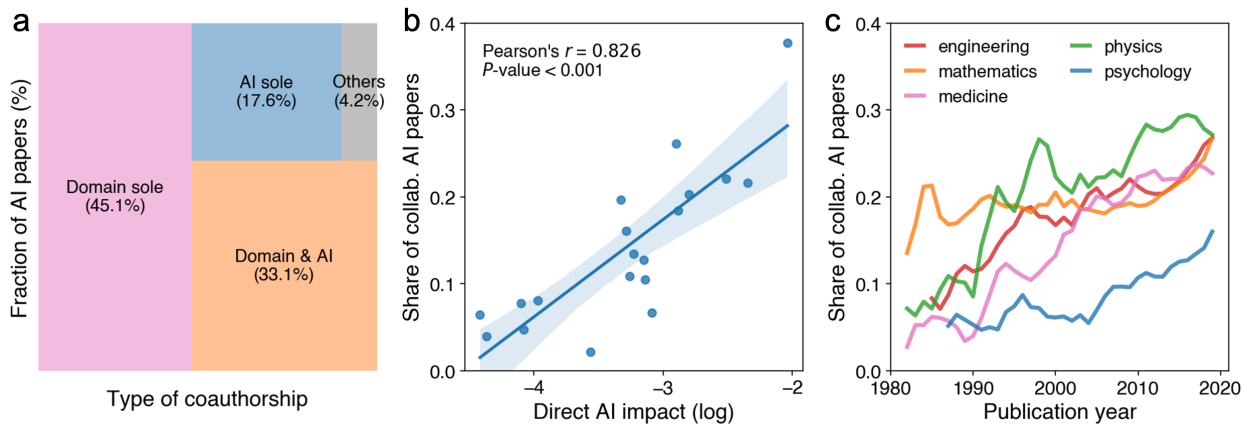

**Supplementary Figure 15. Robustness test of results for identifying primary AI authors. (a)** The treemap chart shows the fraction of AI papers. **(b)** The positive correlation between AI impact and the share of collaborative AI papers. The correlation was determined using a two-sided Pearson's correlation test. Linear fit (centre lines) of the data with 95% confidence intervals (error bands) is shown. **(c)** The increasing trends in the share of collaborative AI papers in major scientific disciplines.

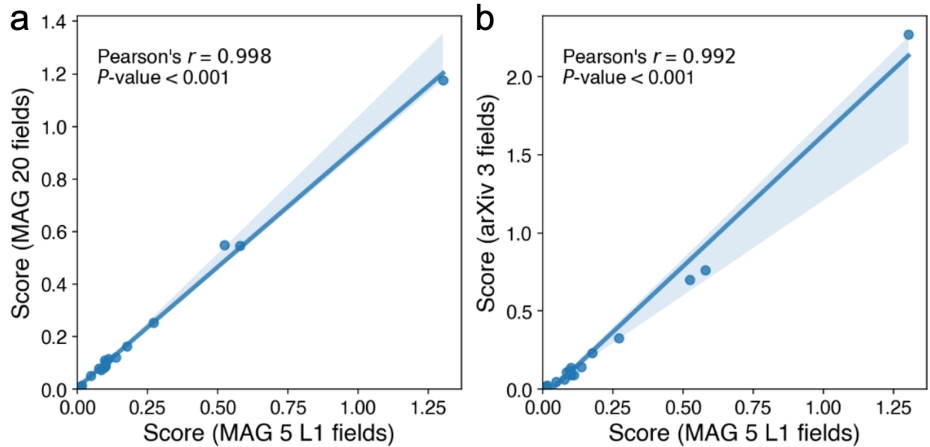

**Supplementary Figure 16. Robustness tests of the result using two alternative definitions of AI-related papers. (a)** Identify AI research using the top 20 most frequent AI n-grams. **(b)** Identify AI research using four arXiv subject categories. The direct AI impact scores of research disciplines under different definitions are highly correlated with each other. Throughout all panels, the correlations were determined using a two-sided Pearson's correlation test, and linear fits (centre lines) of the data with 95% confidence intervals (error bands) are shown.

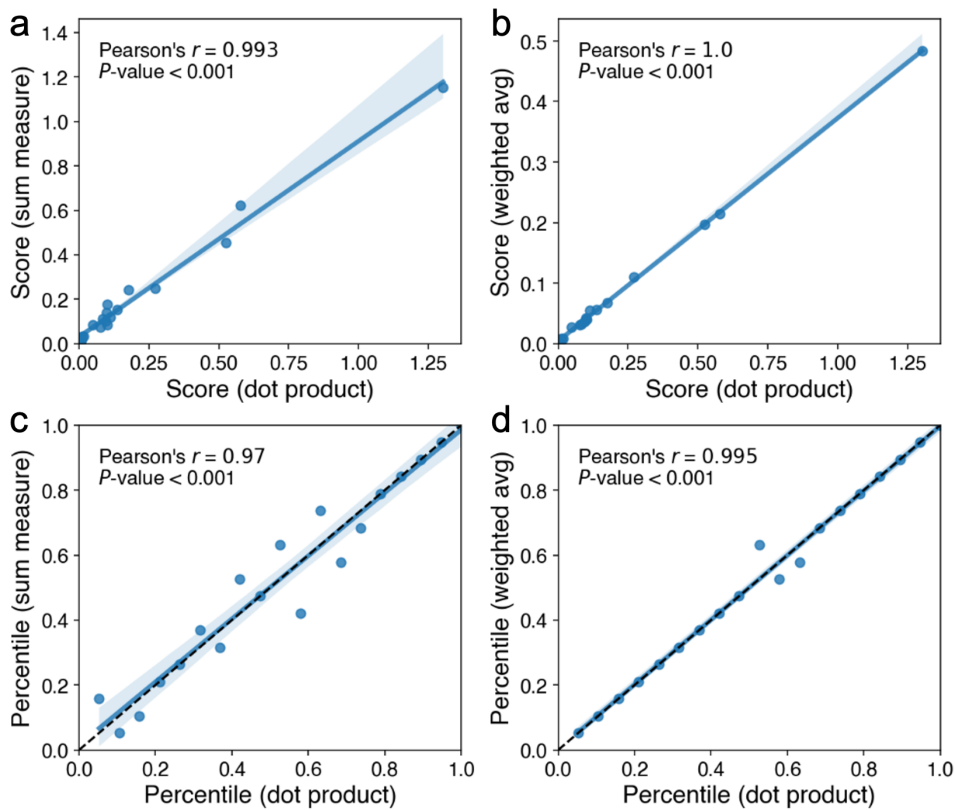

**Supplementary Figure 17. The high correlation between direct AI impact scores.** The scores for disciplines are calculated using different measures. **(a)** Comparing the scores based on the dot product measure and the sum measure. **(b)** Comparing the scores based on the dot product measure and the weighted average measure. **(c)** Comparing the percentiles of raw values based on the dot product measure and the sum measure. **(d)** Comparing the percentiles of raw values based on the dot product measure and the weighted average measure. Throughout all panels, the correlation determined using a two-sided Pearson's correlation test, and linear fits (centre lines) of the data with 95% confidence intervals (error bands) are shown.

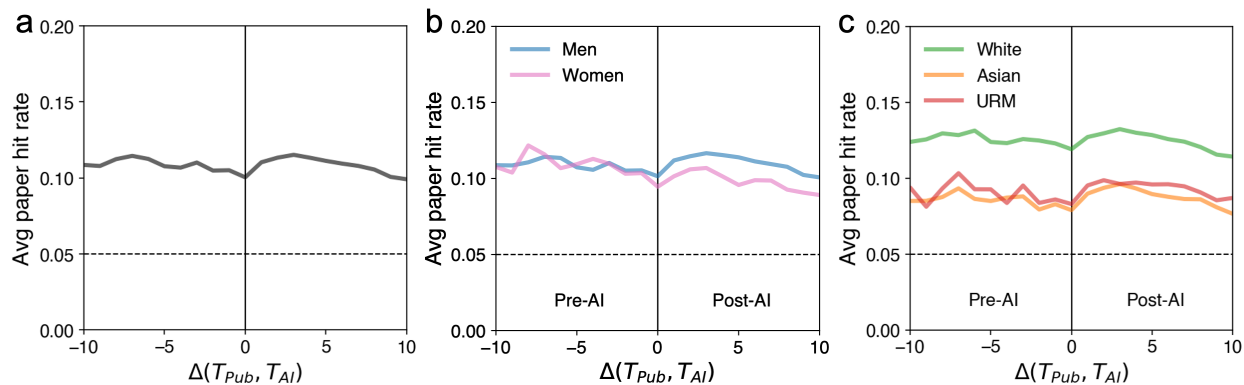

**Supplementary Figure 18. The average hit rate of papers published by authors before and after engaging in AI research. (a)** The results for authors across all demographic groups. **(b)** The results for men or women authors. **(c)** The results for white, Asian, or URM authors. Here only AI researchers are considered. The x-axis is the time difference between the publication year and the starting year of each author's AI engagement.

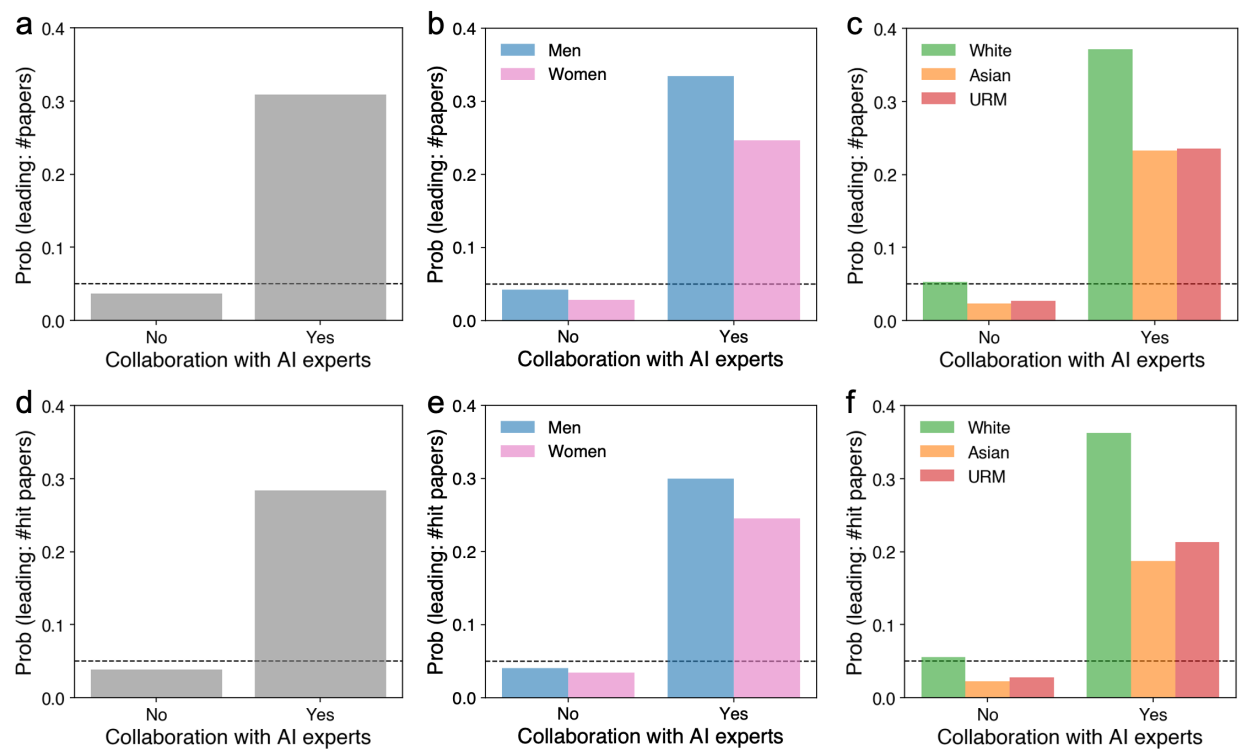

**Supplementary Figure 19. The probability of authors leading the way in their field is based on two metrics. (a-c)** Their number of papers. **(d-f)** Their number of hit papers. Panels (a) and (d) show the results for all demographic groups. Panels (b) and (e) show the results for gender, and Panels (c) and (f) show the results for race.

# Supplementary Tables

**Supplementary Table 1. Alignment between direct and potential AI impact.** The top three subfields within each discipline by the direct AI impact score and the potential AI impact score.

| No. | Discipline | Top 3 subfields (direct AI impact) | Top 3 subfields (potential AI impact) |
|---|---|---|---|
| 1 | computer science | pattern recognition, machine learning, computer vision | pattern recognition, computer vision, machine learning |
| 2 | mathematics | algorithm, mathematical optimization, statistics | algorithm, statistics, computational science |
| 3 | engineering | computer engineering, industrial engineering, control engineering | computer engineering, control engineering, embedded system |
| 4 | geology | remote sensing, soil science, mineralogy | remote sensing, geodesy, petroleum engineering |
| 5 | physics | meteorology, acoustics, medical physics | medical physics, acoustics, meteorology |
| 6 | geography | remote sensing, cartography, meteorology | remote sensing, geodesy, cartography |
| 7 | environmental science | soil science, water resource management, environmental engineering | environmental engineering, soil science, water resource management |
| 8 | psychology | cognitive science, cognitive psychology, communication | mathematics education, cognitive psychology, applied psychology |
| 9 | chemistry | mineralogy, chromatography, environmental chemistry | mineralogy, analytical chemistry, food science |
| 10 | economics | econometrics, management science, actuarial science | operations management, econometrics, environmental economics |
| 11 | biology | biological system, computational biology, neuroscience | biological system, computational biology, animal science |
| 12 | medicine | nuclear medicine, optometry, medical physics | nuclear medicine, medical physics, radiology |
| 13 | materials science | composite material, metallurgy, optoelectronics | optoelectronics, nanotechnology, composite material |
| 14 | business | risk analysis, actuarial science, process management | process management, risk analysis, actuarial science |
| 15 | sociology | communication, demography, regional science | regional science, communication, demography |
| 16 | philosophy | linguistics, humanities, epistemology | linguistics, humanities, epistemology |
| 17 | political science | public relations, law, public administration | public relations, public administration, law |
| 18 | art | visual arts, humanities, aesthetics | humanities, visual arts, aesthetics |
| 19 | history | archaeology, genealogy, ethnology | archaeology, genealogy, ethnology |

**Supplementary Table 2. The top 20 most frequent AI n-grams based on AI publication in MAG.** These AI n-grams can be matched to the MAG "field of study" categories from the L1 level to the L5 level.

| No. | AI n-grams (MAG field categories) | No. | AI n-grams (MAG field categories) |
|---|---|---|---|
| 1 | machine learning | 11 | recurrent neural network |
| 2 | convolutional neural network | 12 | decision tree |
| 3 | deep learning | 13 | reinforcement learning |
| 4 | support vector machine | 14 | supervised learning |
| 5 | deep neural networks | 15 | pattern recognition |
| 6 | artificial intelligence | 16 | natural language processing |
| 7 | computer vision | 17 | genetic algorithm |
| 8 | random forest | 18 | machine translation |
| 9 | artificial neural network | 19 | word embedding |
| 10 | generative adversarial network | 20 | extreme learning machine |

## Supplementary References

1. Sinha, A., et al., *An overview of Microsoft Academic Service (MAS) and applications*, in *Proceedings of the 24th International Conference on World Wide Web*. 2015, WWW: Florence, Italy. p. 243-246.
2. Wang, K.S., et al., *A review of microsoft academic services for science of science studies.* Frontiers in Big Data, 2019. **2**.
3. Wang, K.S., et al., *Microsoft Academic Graph: When experts are not enough.* Quantitative Science Studies, 2020. **1**(1): p. 396-413.
4. Palayew, A., et al., *Pandemic publishing poses a new COVID-19 challenge.* Nature Human Behaviour, 2020. **4**(7): p. 666-669.
5. Viglione, G., *Are women publishing less during the pandemic? Here's what the data say.* Nature, 2020. **581**(7809): p. 365-366.
6. Raynaud, M., et al., *Impact of the COVID-19 pandemic on publication dynamics and non-COVID-19 research production.* BMC Medical Research Methodology, 2021. **21**(1): p. 255.
7. Gao, J., et al., *Potentially long-lasting effects of the pandemic on scientists.* Nature Communications, 2021. **12**(1).
8. Singh Chawla, D., *Massive open index of scholarly papers launches.* Nature, 2022.
9. Priem, J., H. Piwowar, and R. Orr, *OpenAlex: A fully-open index of scholarly works, authors, venues, institutions, and concepts.* arXiv preprint arXiv:2205.01833, 2022.
10. Lin, Z., et al., *SciSciNet: A large-scale open data lake for the science of science research.* Scientific Data, 2023.
11. Microsoft Academic. *Microsoft Academic Graph*. 2021; Available from: https://zenodo.org/record/6511057.
12. Microsoft Academic. *Expanding Concept Understanding in Microsoft Academic Graph*. 2020; Available from: https://www.microsoft.com/en-us/research/project/academic/articles/expanding-concept-understanding-in-microsoft-academic-graph/.
13. Shen, Z., H. Ma, and K. Wang, *A web-scale system for scientific knowledge exploration*, in *Proceedings of Association for Computational Linguistics*. 2018: Melbourne, VIC. p. 87-92.
14. PatentsView. *USPTO PatentsView*. 2019; Available from: http://www.patentsview.org.
15. Toole, A., C. Jones, and S. Madhavan, *Patentsview: An open data platform to advance science and technology policy.* 2021, USPTO Economic Working Paper No. 2021-1.
16. USPTO. *USPTO Research datasets*. 2020; Available from: https://www.uspto.gov/ip-policy/economic-research/research-datasets.
17. Nowogrodzki, A., *Mining the Secrets of College Syllabuses.* Nature, 2016. **539**(7627): p. 125-126.
18. Chau, H., et al., *Connecting higher education to workplace activities and earnings.* 2023. **18**(3): p. e0282323.
19. Börner, K., et al., *Skill discrepancies between research, education, and jobs reveal the critical need to supply soft skills for the data economy.* Proceedings of the National Academy of Sciences, U.S.A., 2018. **115**(50): p. 12630-12637.
20. National Center for Science and Engineering Statistics. *Survey of Doctorate Recipients*. 2019; Available from: https://www.nsf.gov/statistics/srvydoctoratework/.
21. Huang, J., et al., *Historical comparison of gender inequality in scientific careers across countries and disciplines.* Proceedings of the National Academy of Sciences, U.S.A., 2020. **117**(9): p. 4609-4616.

22. Kozlowski, D., et al., *Avoiding bias when inferring race using name-based approaches.* Plos One, 2022. **17**(3).
23. Santamaria, L. and H. Mihaljevic, *Comparison and benchmark of name-to-gender inference services.* Peerj Computer Science, 2018.
24. NCSES. *2021 Survey of Doctorate Recipients*. 2021; Available from: https://www.nsfsdr.org/.
25. Bloom, N., et al., *The diffusion of disruptive technologies.* NBER Working Paper w28999, 2021.
26. Acemoglu, D., et al., *AI and jobs: Evidence from online vacancies.* NBER Working Paper w28257, 2020.
27. Bianchini, S., M. Muller, and P. Pelletier, *Artificial intelligence in science: An emerging general method of invention.* Research Policy, 2022. **51**(10).
28. Littmann, M., et al., *Validity of machine learning in biology and medicine increased through collaborations across fields of expertise.* Nature Machine Intelligence, 2020. **2**(1): p. 18-24.
29. Cockburn, I.M., R. Henderson, and S. Stern, *The impact of artificial intelligence on innovation: An exploratory analysis*, in *The Economics of Artificial Intelligence: An Agenda*. 2018, University of Chicago Press: Chicago, IL. p. 115-146.
30. Klinger, J., J. Mateos-Garcia, and K. Stathoulopoulos, *A narrowing of AI research?* arXiv preprint arXiv:2009.10385, 2020.
31. Tang, X., et al., *The pace of artificial intelligence innovations: Speed, talent, and trial-and-error.* Journal of Informetrics, 2020. **14**(4): p. 101094.
32. Krenn, M., et al., *Forecasting the future of artificial intelligence with machine learning-based link prediction in an exponentially growing knowledge network.* Nature Machine Intelligence, 2023. **5**: p. 1326-1335.
33. Klinger, J., J. Mateos-Garcia, and K. Stathoulopoulos, *Deep learning, deep change? Mapping the development of the Artificial Intelligence General Purpose Technology.* arXiv preprint arXiv:1808.06355, 2018.
34. Frank, M.R., et al., *The evolution of citation graphs in artificial intelligence research.* Nature Machine Intelligence, 2019. **1**(2): p. 79-85.
35. AlShebli, B., et al., *China and the US produce more impactful AI research when collaborating together.* arXiv preprint arXiv:2304.11123, 2023.
36. AlShebli, B., et al., *Beijing's central role in global artificial intelligence research.* Scientific Reports, 2022. **12**(1).
37. Nivre, J. and J. Nilsson, *Pseudo-projective dependency*, in *Proc. of the 43rd Annual Meeting of the Association for Computational Linguistics (ACL)*. 2005, ACL. p. 99-106.
38. Honnibal, M. and M. Johnson, *An improved non-monotonic transition system for dependency parsing*, in *Proceedings of the 2015 Conference on Empirical Methods in Natural Language Processing*. 2015, Association for Computational Linguistics. p. 1373-1378.
39. Benetka, J.R., J. Krumm, and P.N. Bennett, *Understanding context for tasks and activities*, in *Proceedings of the 2019 Conference on Human Information Interaction and Retrieval*. 2019, ACM. p. 133-142.
40. Fujii, H. and S. Managi, *Trends and priority shifts in artificial intelligence technology invention: A global patent analysis.* Economic Analysis and Policy, 2018. **58**: p. 60-69.
41. Giczy, A.V., N.A. Pairolero, and A.A. Toole, *Identifying artificial intelligence (AI) invention: A novel AI patent dataset.* Journal of Technology Transfer, 2022. **47**(2): p. 476-505.

42. Santos, R.S. and L.L. Qin, *Risk Capital and Emerging Technologies: Innovation and Investment Patterns Based on Artificial Intelligence Patent Data Analysis.* Journal of Risk and Financial Management, 2019. **12**(4).
43. Verendel, V., *Tracking artificial intelligence in climate inventions with patent data.* Nature Climate Change, 2023. **13**: p. 40-47.
44. WIPO, *PATENTSCOPE Artificial Intelligence Index*. 2020.
45. WIPO, *WIPO Technology Trends 2019 – Artificial Intelligence*. 2019, World Intellectual Property Organization: Geneva.
46. Miric, M., N. Jia, and K. Huang, *Using supervised machine learning for large-scale classification in management research: The case for identifying artificial intelligence patents.* Strategic Management Journal, 2023. **44**(2): p. 491-519.
47. Webb, M., *The impact of artificial intelligence on the labor market.* SSRN 3482150, 2019.
48. Felten, E.W., M. Raj, and R. Seamans, *A method to link advances in artificial intelligence to occupational abilities.* AEA Papers and Proceedings, 2018. **108**: p. 54-57.
49. Felten, E., M. Raj, and R. Seamans, *Occupational, industry, and geographic exposure to artificial intelligence: A novel dataset and its potential uses.* Strategic Management Journal, 2021. **42**(12): p. 2195-2217.
50. Brynjolfsson, E. and T. Mitchell, *What can machine learning do? Workforce implications.* Science, 2017. **358**(6370): p. 1530-1534.
51. Brynjolfsson, E., T. Mitchell, and D. Rock, *What can machines learn, and what does it mean for occupations and the economy?* AEA Papers and Proceedings, 2018. **108**: p. 43-47.
52. Kogan, L., et al., *Technology, vintage-specific human capital, and labor displacement: Evidence from linking patents with occupations*. 2022, NBER Working Paper No. w29552.
53. Frank, M., Y.Y. Ahn, and E. Moro, *AI exposure predicts unemployment risk*, in *arXiv preprint arXiv:2308.02624*. 2023.
54. Atalay, E., et al., *The evolution of work in the United States.* American Economic Journal: Applied Economics, 2020. **12**(2): p. 1-34.
55. Kiperwasser, E. and Y. Goldberg, *Simple and accurate dependency parsing using bidirectional LSTM feature representations.* Transactions of the Association for Computational Linguistics, 2016. **4**: p. 313-327.
56. Kim, I.C. and Y.G. Jung, *Using Bayesian networks to analyze medical data*, in *Proceedings of the Third International Workshop on Machine Learning and Data Mining in Pattern Recognition*. 2003: Leipzig, Germany. p. 317-327.
57. Leskovec, J., A. Rajaraman, and J.D. Ullman, *Mining of massive data sets*. 2020, New York, NY: Cambridge University Press.
58. Wu, L., L. Hitt, and B.W. Lou, *Data analytics, innovation, and firm productivity.* Management Science, 2020. **66**(5): p. 2017-2039.
59. OSP. *Open Syllabus Dataset Documentation*. 2020; Available from: https://docs.opensyllabus.org/index.html.
60. Malitz, G.S., *A classification of instructional programs (CIP)*. 1987, Washington, D.C.: Center for Education Statistics, Office of Educational Research and Improvement, US Department of Education.
61. Cook, K., et al., *The future of work in black america*. 2019, McKinsey Insights.
62. Autor, D., *The faltering escalator of urban opportunity*, in *Securing Our Economic Future*. 2020, The Aspen Institute: Washington, DC, USA. p. 108-136.

63. Cornell University Library. *arXiv Dataset: arXiv dataset and metadata of 1.7M+ scholarly papers across STEM*. 2023; Available from: https://www.kaggle.com/datasets/Cornell-University/arxiv/data.
64. Sourati, J. and J. Evans, *Complementary artificial intelligence designed to augment human discovery.* arXiv preprint arXiv:2207.00902, 2022.
65. Baruffaldi, S., et al., *Identifying and measuring developments in artificial intelligence: Making the impossible possible*, in *OECD Science, Technology and Industry Working Papers No. 2020/05*. 2020.
66. Van Buskirk, I. *Name-Based Gender Classification*. 2021; Available from: https://osf.io/tz38q/.
67. Van Buskirk, I., A. Clauset, and D.B. Larremore, *An open-source cultural consensus approach to name-based gender classification*, in *Proceedings of the International AAAI Conference on Web and Social Media*. 2023. p. 866-877.
68. Liang, L. and D.E. Acuna. *demographicx: A Python package for estimating gender and ethnicity using deep learning transformers*. 2021; Available from: https://doi.org/10.5281/zenodo.4898367.
69. Ke, Q., et al., *A dataset of mentorship in bioscience with semantic and demographic estimations.* Scientific Data, 2022. **9**(1): p. 467.